%% file: arxiv.tex
\documentclass[aps,twocolumn,a4paper,amssymb,amsmath,prl,reprint,showpacs,groupedaddress]{revtex4-1}
\usepackage{amsmath,amssymb}
\usepackage[hiresbb,pdftex]{graphicx}		
\usepackage{bm,braket,latexsym,multirow,longtable}
\usepackage[pdftex]{xcolor}

\usepackage[pdftex,
draft=false,
bookmarks=true,
bookmarksnumbered=false,
bookmarksopen=false,
bookmarksopenlevel=4,
colorlinks=true,
anchorcolor=black,
citecolor=blue,
filecolor=magenta,
linkcolor=blue,
linkbordercolor={1 0 0},
urlcolor=violet,
pdfborder={0 0 1},
pdftitle={},
pdfauthor={},
pdfsubject={},
pdfkeywords={},
pdfpagemode=UseThumbs,
]{hyperref}

\newcommand{\mr}[2]{#1 \, {\mathrm{ \,  #2 \,}}}
\newcommand{\aizu}[2]{ #1 \, \textit{F} \, #2}

\begin{document}

\title{Symmetry analysis of current-induced switching of antiferromagnets}
\author{Hikaru Watanabe}
\email[]{watanabe.hikaru.43n@st.kyoto-u.ac.jp}
\affiliation{Department of Physics, Graduate School of Science, Kyoto University, Kyoto 606-8502, Japan}
\author{Youichi Yanase}
\affiliation{Department of Physics, Graduate School of Science, Kyoto University, Kyoto 606-8502, Japan}
\date{\today}

\begin{abstract}
Antiferromagnets are robust  to external electric and magnetic fields, and hence are seemingly uncontrollable. Recent studies, however, realized the electrical manipulations of antiferromagnets by virtue of the antiferromagnetic Edelstein effect. We present a general symmetry analysis of electrically switchable antiferromagnets based on group-theoretical approaches. Furthermore, we identify a direct relation between switchable antiferromagnets and the ferrotoroidic order. The concept of the ferrotoroidic order clarifies the unidirectional nature of switchable antiferromagnets and provides a criterion for the controllability of antiferromagnets. The scheme paves a way for perfect writing and reading of switchable antiferromagnets.
\end{abstract}
\maketitle

The spin degree of freedom is highly controllable and has opened a new paradigm of electronics~\cite{Zutic2004,Tserkovnyak2005,Chappert2007a,Kent2015}. Especially, spin manipulation by an interplay with other degrees of freedom such as charge and valley is attracting much interest in the field of spintronics. The concept is widespread in condensed matter physics, \textit{e.g.,} superconductors~\cite{Linder2015} and topological materials~\cite{Shiomi2014}.

Recently, the possibility of manipulating antiferromagnets has been recognized~\cite{MacDonald2011,Gomonay2014a,Jungwirth2016,Baltz2018,Manchon2018currentinduced}, while most spintronics devices are based on ferromagnets. The antiferromagnet, which in itself has neither electric nor magnetic polarizations, is robust to external electric and magnetic fields, in contrast to ferroelectric or ferromagnetic material-based devices. Hence, antiferromagnets are considered to be a new candidate for a nonvolatile memory device~\cite{Wadley2016}. 

The switching of antiferromagnets has been explored in various ways: by using a spin-transfer torque or spin current in heterostructures~\cite{Gomonay2010SpinTransfer,Moriyama2018Cr2O3,Moriyama2018spintorqueswitch} or magnetoelectric effects in bulk antiferromagnets~\cite{Kosub2017}. In particular, manipulation by an electric current has significantly promoted the controllability of antiferromagnets. The switching mechanism utilizes current-induced antiferroic magnetization, namely, the antiferromagnetic (AFM) Edelstein effect. Following a theoretical proposal submitted independently by some groups~\cite{Yanase2014,Zelezny2014}, experimental realizations of switching have been achieved in CuMnAs~\cite{Wadley2016} and Mn$_2$Au~\cite{Bodnar2018}.

The key to the AFM Edelstein effect is a \textit{locally} noncentrosymmetric crystalline structure, in which the local symmetry of certain sites has no parity symmetry in spite of a globally centrosymmetric crystalline symmetry. The sublattice degree of freedom plays an important role in such systems. The parity symmetry is preserved since the atoms in each sublattice are interchanged by the parity operation. Then, spin-momentum locking arises in a sublattice-dependent manner although uniform spin-momentum locking is forbidden due to the globally centrosymmetric crystal symmetry~\cite{FischerSuperconductivity2011,Zhang2014c,Ciccarelli2015,frigeriThesis,MaruyamaMultilayer2012}. Accordingly, nonequilibrium \textit{antiferroic} spin polarization is induced under electric current. This is an analog to the Edelstein effect, that is, current-induced ferroic spin polarization~\cite{Edelstein1990,Garate2009,Manchon2008,Manchon2009}.

The switchable AFM order, which shares the same symmetry with current-induced antiferroic magnetization, breaks both of inversion and time-reversal symmetry although it preserves translational symmetry. It is noteworthy that the combined symmetry of parity and time-reversal operations, namely, $\mathcal{PT}$ symmetry, is preserved. The $\mathcal{PT}$ symmetry forbids electric polarization and net magnetization and ensures invulnerability to external electric or magnetic fields.

The properties of the AFM Edelstein effect have been investigated in previous studies~\cite{Yanase2014,Zelezny2014,Hayami2014b,Zelezny2017,Watanabe2017magnetic}. The general criterion for determining electrically switchable antiferromagnets, however, remains unclear. An incomplete understanding of switchable antiferromagnets has disrupted further explorations of candidate materials except for the existing candidates~\cite{Wadley2016,Bodnar2018}.

In our work, we present a general criterion for the current-induced switching of antiferromagnets. A symmetry analysis based on the magnetic representation theory and the Aizu species clarifies what kind of AFM order can be manipulated by the AFM Edelstein effect. Furthermore, the analysis links switchable antiferromagnets and ferrotoroidic order. Thus, this work not only identifies many candidate materials but also provides a clear viewpoint of AFM spintronics. In the following, we do not discuss the effect of spin-transfer torque and spin current, since we focus on bulk antiferromagnets which are insensitive to surfaces.

\textit{Representation analysis. }We first present a symmetry analysis of magnetic modes induced by the AFM Edelstein effect. The analysis is carried out with the use of a magnetic representation theory~\cite{inui1990group,Bertaut1968,izyumov1991neutron,supplemental}. We focus on centrosymmetric systems where the $\mathcal{PT}$ symmetry is preserved, though it is straightforward to extend the analysis to noncentrosymmetric systems. Noncentrosymmetric magnets (Ga,Mn)As and MnSiN$_2$ are exemplified in Supplemental Material~\cite{supplemental}. 

The magnetic modes realized by the AFM Edelstein effect  do not lead to any translational symmetry breaking, and hence they are characterized by the N\'eel vector $\bm{Q}=\bm{0}$. The formula of the AFM Edelstein effect is written as
  \begin{equation}
    \hat{m}^\mathrm{AF} =\hat{\kappa}~\bm{j}, \label{antiferromagnetic_Edelstein}
  \end{equation}
where the susceptibility tensor $\hat{\kappa}$ has symmetry determined by the crystalline structure~\cite{Ciccarelli2015,Zelezny2017}. Therefore, supposing systems in the paramagnetic phase, we here identify the AFM mode $\hat{m}^\mathrm{AF}$ and investigate which components of $\hat{\kappa}$ are allowed.

The allowed magnetic symmetry with  $\bm{Q}=\bm{0}$ is described by a magnetic representation
  \begin{equation}
  	\Gamma_{\bm{G}}^\textrm{mag} \left(\bm{H} \right)= \Gamma_{\bm{G}}^\mathrm{P} (\bm{H}) \otimes \Gamma_{\bm{G}}^{\bm{M}},\label{magnetic_representation}
  \end{equation}
where $\bm{G}$ and $\bm{H}$ denote a crystal group and site-symmetry group of magnetic sites, respectively. $\Gamma_{\bm{G}}^\mathrm{P} $ is a permutation representation of magnetic sites and $\Gamma_{\bm{G}}^{\bm{M}}$ is a representation of an axial vector. The basis $\hat{m}^\textrm{AF}$ of allowed magnetic modes is explicitly denoted as $m_{\mu}^{(\tau)}$, where $\tau$ and $\mu$ are indices of the basis of $\Gamma_{\bm{G}}^\mathrm{P} $ and $\Gamma_{\bm{G}}^{\bm{M}} $, respectively.

The response formula~\eqref{antiferromagnetic_Edelstein} is explicitly recast
    \begin{equation}
    m_{\mu}^{(\tau)} = \kappa_{\tau \mu;\nu} j_{\nu}. \label{antiferromagnetic_Edelstein_with_coefficients}
    \end{equation}
The coefficient $\kappa_{\tau \mu;\nu}$ is transformed by a symmetry operation $g \in \bm{G}$ as~\cite{Seemann2015}
      \begin{equation}
      g \left(  \kappa_{\tau \mu;\nu} \right)  = \sum_{\rho, \lambda,\kappa} \kappa_{\rho \lambda;\kappa} [ D^\textrm{(P)} (g)]_{\rho \tau} [D^{(\bm{M})}(g)]_{\lambda \mu} [D^{(\bm{j})}(g)]_{\kappa \nu}, \label{antiferromagnetic_Edelstein_transformation}
      \end{equation}
where $D^\textrm{(P)}$, $D^{(\bm{M})}$, and $D^{(\bm{j})}$ are representation matrices of the sublattice permutation, axial vector, and polar vector, respectively. According to Neumann's principle, the transformed susceptibility tensor should satisfy $ g \left(  \hat{\kappa} \right) = \hat{\kappa}$~\cite{birss1964symmetry,cracknell2016magnetism}. Thus, the susceptibility tensor $\hat{\kappa}$ is subject to constraints from the crystal group $\bm{G}$.

An algebraic calculation by Eq.~\eqref{antiferromagnetic_Edelstein_transformation} identifies the symmetry-adapted form of $\kappa_{\tau \mu;\nu}$. To examine the symmetry constraints between magnetic structures and electric currents, it is practical to decompose the representation of the susceptibility tensor $\kappa_{\tau \mu;\nu}$ into irreducible representations of $\bm{G}$, that is, $\{ \Gamma_{\bm{G}}^{(\alpha)} \}$. The decomposition is obtained as
  \begin{equation}
  \Gamma_{\bm{G}}^\mathrm{mag} \left(\bm{H} \right) \otimes \Gamma_{\bm{G}}^{\bm{j}} = \sum_{\alpha} q_\alpha \Gamma_{\bm{G}}^{(\alpha)},\label{edelstein_decomposition}
  \end{equation}
where a coefficient $q_\alpha$ denotes a frequency of $ \Gamma_{\bm{G}}^{(\alpha)} $ in the summation. $\Gamma_{\bm{G}}^{\bm{j}}$ is a representation of the polar vector. The coefficient $q_1$ for the identity representation $\Gamma_{\bm{G}}^{(1)}$ gives the number of independent components of $\hat{\kappa}$.

Here, we summarize the symmetry constraints for switchable antiferromagnets. Each irreducible representation $\Gamma_{\bm{G}}^{(\alpha)}$ has inversion parity, since the crystal group $\bm{G}$ is centrosymmetric. Although both $\Gamma_{\bm{G}}^{\bm{M}}$ and $\Gamma_{\bm{G}}^{\bm{j}}$ are representations of vector quantities, they have opposite parity. Thus, the permutation representation $\Gamma_{\bm{G}}^\mathrm{P} $ should comprise odd-parity irreducible representations to satisfy $q_1 \neq 0$. This means that a locally noncentrosymmetric property of magnetic sites is required for the AFM Edelstein effect as mentioned in previous studies~\cite{Yanase2014,Zelezny2014,Zelezny2017}. Furthermore, the magnetic representation $\Gamma_{\bm{G}}^\textrm{mag}$ should comprise a polar representation $\Gamma_{\bm{G}}^{\bm{j}}$. Owing to the time-reversal even/odd ($\mathcal{T}$ even/$\mathcal{T}$ odd) nature of the representation $\Gamma_{\bm{G}}^\mathrm{P}$/ $\Gamma_{\bm{G}}^{\bm{M}}$, the magnetic mode $m_{\mu}^{(\tau)}$ is $\mathcal{T}$ odd and leads to the time-reversal symmetry breaking.

From the above analysis we conclude that the current-induced magnetic structure is polar and  magnetic ($\mathcal{T}$ odd). It follows that the switchable AFM order by the AFM Edelstein effect contains a \textit{toroidal moment} $\bm{T}$~\cite{Spaldin2008}. We stress that the N\'eel vector is $\bm{Q}=\bm{0}$. Thus, all the switchable AFM order is regarded as a ferroic toroidal order, namely, ferrotoroidic order. This is a criterion of materials for AFM spintronics.

The toroidic nature of switchable antiferromagnets is intuitively understood by the fact that the electric current gives rise to a shift of the Fermi surface and produces ``polarization'' in momentum space. Such polarization shares the same symmetry with the toroidal moment as we have shown in the group-theoretical classification~\cite{Watanabe2018grouptheoretical}.

\textit{Aizu species. }Regarding the switchable AFM order as ferrotoroidic order, we may clarify the possibility of AFM domain switching by making use of the \textit{Aizu species}.

In general, a phase transition reduces the symmetry operations of a disordered phase. The symmetry relation between the disordered and ordered phases is formulated by a group-theoretical method. By supposing the crystal group $\bm{G}\left( \bm{K}\right)$ in the disordered (ordered) phase, the coset decomposition of $\bm{G}$ by $\bm{K}$ is obtained as
    \begin{equation}
      \bm{G} = g_1 \bm{K} +g_2 \bm{K}+\cdots g_N \bm{K}, \label{GK_decomposition}
    \end{equation}  
where $g_1 \in \bm{K}$ and $g_j \not \in \bm{K} \left(j\neq 1 \right)$. $N$ is the order of $\bm{G}$ divided by that of $\bm{K}$. A domain state $s_1$, which is invariant to the symmetry operations of $\bm{K}$, is transformed into other domain states by symmetry operations of $ g_j \bm{K} \left(j\neq 1 \right)$. Therefore, the coset decomposition~\eqref{GK_decomposition} shows the relation between domain states $\{ s_j \}$,
  \begin{equation}
    s_{j} = g_j g_i^{-1} s_{i}, \label{domains_algebraic_relation}
  \end{equation}
where the domain $s_j$ is invariant to the symmetry operations of $\bm{K}_j =g_j \bm{K} g_j^{-1}$. 

Domain properties of the ordered phase are classified by the Aizu species~\cite{Aizu1966,Aizu1969,Aizu1970a,Schmid1999,Schmid2008,Litvin2008,Hlinka2016a}, the ensemble of pairs of $\bm{G}$ and $\bm{K}$ written as $\aizu{\bm{G}}{\bm{K}}$. In the Aizu species classification, the species $\aizu{\bm{G}}{\bm{K}}$ is characterized by physical quantities such as electric polarization, magnetization, strain, and toroidal moment. In the case of ferrotoroidic order, we first assign a domain $s_1$ with a toroidal moment $\bm{T}^{(1)}$. Correspondingly, we obtain the toroidal moment
  \begin{equation}
    \bm{T}^{(j)} = g_j \bm{T}^{(1)}
  \end{equation}
for another domain $s_j$. The number of possible toroidal moments is determined by a given species, since the species $\aizu{\bm{G}}{\bm{K}}$ imposes the algebraic relation between domains as Eq.~\eqref{domains_algebraic_relation}. Therefore, the Aizu species $\aizu{\bm{G}}{\bm{K}}$ is classified as full/partial/zero toroidic, when the domain states are completely/partially/not distinguishable by the toroidal moment $\bm{T}$. The classification is summarized in Table~\ref{aizu_classify}.
    \begin{table}[htbp]
      \caption{The classification of Aizu species ${\protect \aizu{\bm{G}}{\bm{K}}}$ based on toroidal moments $\{ \bm{T}^{(i)} \}$.}
    \label{aizu_classify}
    \centering
      \begin{tabular}{ll}
      \hline \hline
      $\{ \bm{T}^{(i)} \}$& $\aizu{\bm{G}}{\bm{K}}$ \\ \hline
      $\bm{T}^{(i)}\neq \bm{T}^{(j)}$ for all $i,j$&full toroidic \\ 
      $\bm{T}^{(i)} = \bm{T}^{(j)}$ for some but not all $i,j$   &partial toroidic \\ 
      $\bm{T}^{(i)} = \bm{T}^{(j)} = 0$ for all $i,j$&zero toroidic \\ \hline \hline
      \end{tabular}
    \end{table} 

In a full- or partial toroidic species, the symmetry-adapted field for the toroidal moment $\bm{T}$, that is, electric current $\bm{j}$, energetically distinguishes the domain states completely or partially. The electric current acts on the AFM moment such that  the toroidal moment arising from the AFM mode is aligned along the injected current. Therefore, the classification based on the Aizu species for the ferrotoroidic order clarifies the AFM domains which are controllable by the electric current $\bm{j}$. With a pair of the crystal group $\bm{G}$ and the group for the AFM state $\bm{K}$, the feasibility of the electrical switching of AFM domains is determined by referring to the toroidic property of the Aizu species~\cite{Litvin2008}.

The switchable antiferromagnets should belong to the full- or partial toroidic species. We have identified candidate materials for the switchable AFM order and show a part of the list in Table~\ref{magnetic_candidates}. In Supplemental Material, we can find more candidates and more detailed information~\cite{supplemental}. In the following, we apply our symmetry analysis to some antiferromagnets and reveal the toroidic property of the AFM state.

Similarly,
  \begin{table}[htbp]
  \caption{List of candidate materials. The table lists metallic or semiconducting compounds, crystal point group (PG), direction of toroidal moment ($\bm{T}$) , N\'eel temperatures ($T_\mathrm{N}$), and references (Ref.). More candidates are shown in Supplemental Material~\cite{supplemental}.}
  \label{magnetic_candidates}
  \begin{tabular}{lllll}
    Compounds &PG&$\bm{T}$&$T_\mathrm{N}$&Ref. \\ \hline \hline
  PrMnSbO&$4/mmm$&&$35<T<230$&\cite{kimber2010local}\\
          &&$\{T_x,T_y\}$&35&\cite{kimber2010local}\\
  NdMnAsO&$4/mmm$&&$23<T<359$&\cite{Marcinkova2010,Emery2011b}\\
          &&$\{T_x, T_y\}$&23&\cite{Marcinkova2010,Emery2011b}\\
  DyB$_4$&$4/mmm$&$\{T_x, T_y\}$&$12.7<T<20.3$&{\cite{Fisk1981,Will1979,Ji2007}}\\
  ErB$_4$&$4/mmm$&$\{T_x, T_y\}$&13&{\cite{Fisk1981,Will1979,Will1981}}\\
  Mn$_2$Au&$4/mmm$&$\{T_x, T_y\}$&$>1000$&\cite{Barthem2013}\\
  FeSn$_2$&$4/mmm$&$\{T_x, T_y\}$&$93<T\lesssim378$&\cite{Venturini1987,Armbruster2010}\\
  &&$\{T_x, T_y\}$&$93\lesssim T<378$&\cite{Venturini1987,Armbruster2010}\\
  CuMnAs&$4/mmm$&$\{T_x, T_y\}$&480&\cite{Wadley2013}\\
  U$_3$Ru$_4$Al$_{12}$&$6/mmm$&$T_z$&9.5&\cite{Pasturel2009a,Troc2012}\\
  CaMn$_2$Bi$_2$&$\bar{3}m$&$\{T_x, T_y\}$&154&\cite{Gibson2015}\\
  SrMn$_2$Sb$_2$&$\bar{3}m$&$\{T_x, T_y\}$&110&\cite{Sangeetha2018}\\
  Gd$_5$Ge$_4$&$mmm$&$T_z$&127&\cite{Tan2005,Levin2001} \\
  UCu$_5$In&$mmm$&$T_y$&25&\cite{Tran2001}\\
  YbAl$_{1-x}$Fe$_x$B$_{4}$&$mmm$&$T_x$&&\cite{ybalb4}\\
  \end{tabular}
  \end{table}

\textit{Full toroidic case. }
 As an example of the ferrotoroidic case, we discuss the tetragonal CuMnAs~\cite{Wadley2013} where AFM domain switching has been demonstrated~\cite{Wadley2016}. The crystal group is $4/mmm$ which is represented as $\bm{G}=4/mmm1'$ in magnetic point group notation. The AFM phase is specified by $\bm{K}=mmm'\Braket{x}$, where the symbol $\Braket{x}$ means the twofold rotation symmetry along the $x$ axis. Correspondingly, the Aizu species is denoted by
  \begin{equation}
    \aizu{4/mmm1'}{mmm'\Braket{x}}. \label{CuMnAs_Aizu_species}
  \end{equation}
The coset decomposition~\eqref{GK_decomposition} is obtained as
  \begin{equation}
    \bm{G} = I \bm{K} +P \bm{K} + C_{4z}^+ \bm{K}+  S_{4z}^+ \bm{K}, \label{CuMnAs_decomposition}
  \end{equation}
where $I$, $P$,  and $C_{4z}^+ \left(S_{4z}^+ \right)$ are the identity operation,  the parity operation, and the four-fold (improper) rotation, respectively. The domain $s_1$ with the polar axis $x$ possesses the toroidal moment $\bm{T}^{(1)}= T \hat{x}$. Accordingly, the toroidal moment of the domain $s_2= P s_1$ is obtained as 
  \begin{equation}
  \bm{T}^{(2)} = P \left( T \hat{x}\right) =-T \hat{x}.
  \end{equation} 
Similarly, $\bm{T}^{(3)}=T \hat{y}$ and $\bm{T}^{(4)}=-T \hat{y} $ are obtained for the domains $s_3$ and $s_4$, respectively. Thus, all the domains of CuMnAs have different toroidal moments, and the Aizu species~\eqref{CuMnAs_Aizu_species} is actually full toroidic. Therefore, the AFM state can be completely manipulated by the electric current.

We summarize the properties of the Aizu species in Eq.~\eqref{CuMnAs_Aizu_species} in Table~\ref{CuMnAs_Aizu_species_characterized}. The species~\eqref{CuMnAs_Aizu_species} is zero electric and zero magnetic, and hence AFM domains can hold neither electric polarization $\bm{P}$ nor magnetic polarization $\bm{M}$. These constraints are consistent with the $\mathcal{PT}$ symmetry preserved in the AFM state. On the other hand, the species is partial elastic. It follows that the AFM domains are partially controllable by stress which is the conjugate field to the strain $\hat{\epsilon}$. Thus, the Aizu species analysis is also useful to elucidate the possibility of an indirect switching of the AFM state.

    \begin{table}[htbp]
    \caption{The characterization of the Aizu species of CuMnAs~\cite{Litvin2008}. ``F,'' ``P,'' and ``Z'' represent full, partial, and zero, respectively.}
    \label{CuMnAs_Aizu_species_characterized}
    \centering
    \begin{tabular}{ccccc} \hline \hline 
    \multirow{2}{*}{$\aizu{4/mmm1'}{mmm'\Braket{x}}$}&$\hat{\epsilon}$&$\bm{P}$&$\bm{M}$&$\bm{T}$ \\  
    &P&Z&Z&F\\ \hline\hline
    \end{tabular}
    \end{table}

\begin{figure}[htbp] 
    \centering 
    \includegraphics[width=80mm,clip]{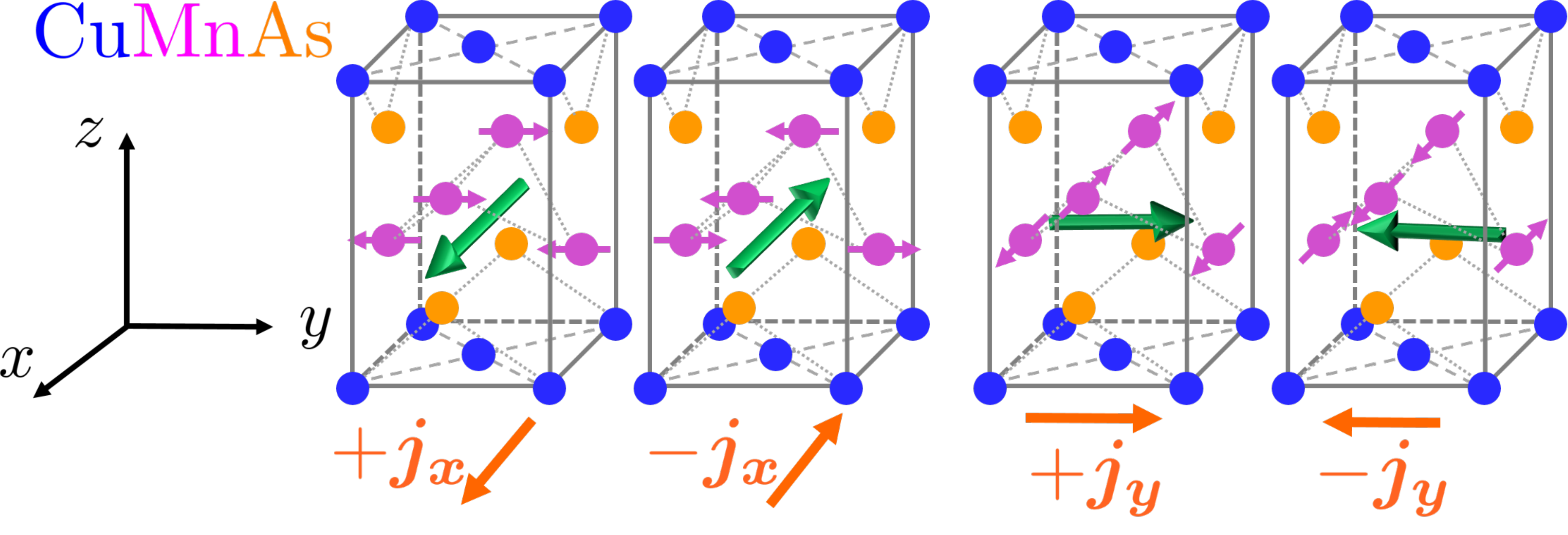} 
    \caption{The correspondence between the AFM domain and the toroidal moment in CuMnAs. The green (purple) colored  arrows represent the toroidal (magnetic) moments. The electric current stabilizing each domain is depicted.
    }\label{CuMnAs_order_toroidal} 
    \end{figure}

To support the Aizu species analysis, we conduct a representation analysis. The magnetic Mn ions are positioned in the crystallographic site with a noncentrosymmetric site-symmetry group $\bm{H}=4mm$, and CuMnAs is locally noncentrosymmetric. The magnetic representation is obtained as 
  \begin{align}
  \Gamma_{\bm{G}}^\textrm{mag}\left(\bm{H} \right)
    &= \Gamma_{\bm{G}}^\mathrm{P} (\bm{H}) \otimes \Gamma_{\bm{G}}^{\bm{M}},\\
    &= A_{2g}+E_{g}+A_{1u}+E_u.\label{CuMnAs_magnetic_representation}
  \end{align}
Then, the product representation~\eqref{edelstein_decomposition} comprises $\Gamma_{\bm{G}}^{(1)}=A_{1g}$, since the $E_u$ mode in Eq.~\eqref{CuMnAs_magnetic_representation} is included in the polar representation $\Gamma_{\bm{G}}^{\bm{j}}= A_{2u}+E_u$. Thus, the AFM Edelstein effect is allowed. The correspondence between the toroidal moment and the AFM order is clarified by the projection operator method~\cite{inui1990group,supplemental}. By the projection operator associated with the basis of $E_u$, the AFM moment aligned along the $\pm x$ axis is revealed to have a toroidal moment $\pm T \hat{y}$. Hence, the in-plane electric current $\pm j_y$ stabilizes the AFM state as shown in Fig~\ref{CuMnAs_order_toroidal}. Similarly, the electric current $\pm j_x$ stabilizes the AFM moment along the $\pm y$ axis which has the toroidal moment $\pm T \hat{x}$. Thus, the representation theory is consistent with the Aizu species analysis.

\textit{Partial toroidic case.}
Next, we discuss a partially controllable AFM state of U$_3$Ru$_4$Al$_{12}$, which belongs to a partial toroidic species.

U$_3$Ru$_4$Al$_{12}$ crystallizes in a hexagonal structure ($\bm{G}= 6/mmm$). The magnetic uranium ions form a kagom\'e lattice~\cite{Pasturel2009a,Troc2012}. Interestingly, the compound shows a compensated and noncollinear AFM order by which the threefold rotation symmetries are broken~\cite{Troc2012}. The Aizu species is given by 
\begin{equation}
	\aizu{6/mmm1'}{mmm'\Braket{z}},
\end{equation}
where $\Braket{z}$ means the twofold rotation symmetry along the $z$ axis. The species is partial toroidic as shown in Table~\ref{U3Ru4Al12_Aizu_species_characterized} and allows the domain states to be partially controllable by the electric current. Following the algebraic relation between the domain states, half of the six domains host the same toroidal moment $\bm{T} \parallel \hat{z}$ which can be inverted by the out-ofplane electric current $j_z$.

    \begin{table}[htbp]
    \caption{The characterization of the Aizu species of U$_3$Ru$_4$Al$_{12}$~\cite{Litvin2008}.}
    \label{U3Ru4Al12_Aizu_species_characterized}
    \centering
    \begin{tabular}{ccccc} \hline \hline 
    \multirow{2}{*}{$\aizu{6/mmm1'}{mmm'\Braket{z}}$}&$\hat{\epsilon}$&$\bm{P}$&$\bm{M}$&$\bm{T}$ \\  
    &P&Z&Z&P\\ \hline\hline
    \end{tabular}
    \end{table}

We also present a representation analysis. The U atoms are positioned in crystallographic sites with the site-symmetry group $\bm{H}=mm2$. The magnetic representation is obtained as
  \begin{align}
  \Gamma_{\bm{G}}^\textrm{mag} \left(\bm{H} \right)
    &= A_{2g}+B_{1g}+B_{2g}+A_{1u}+A_{2u}+B_{1u}\notag \\
    &+2E_{1g}+E_{2g}+E_{1u}+2E_{2u}, \label{U3Ru4Al12_magnetic_representation}
  \end{align} 
which comprises polar representations $A_{2u}$ and $E_{1u}$. The basis of $A_{2u}$ ($E_{1u}$) can be taken as a toroidal moment $\bm{T}\parallel \hat{z}$ $\left( \bm{T} \parallel \{ \hat{x},\hat{y} \} \right)$, and the AFM Edelstein effect is actually allowed when $\bm{j} \parallel \hat{z}$ $\left( \bm{j} \parallel \{ \hat{x},\hat{y} \} \right)$.

\begin{figure}[htbp] 
\centering 
\includegraphics[width=85mm,clip]{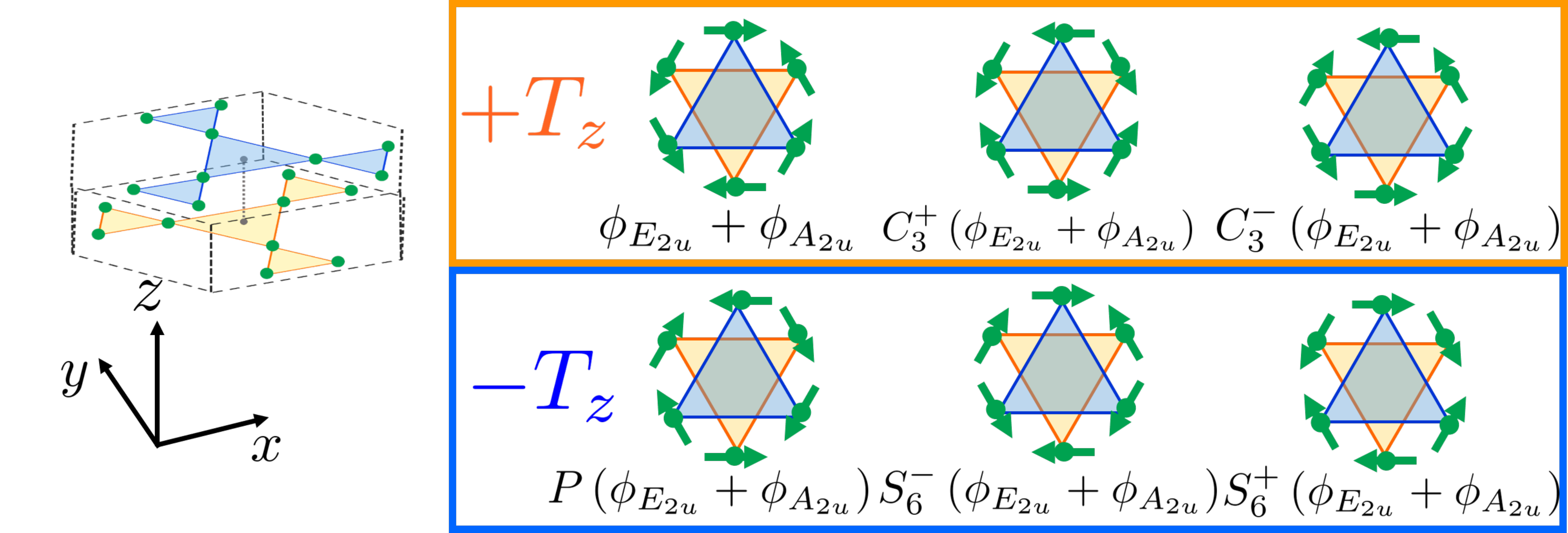}
\caption{The Uranium sites in U$_3$Ru$_4$Al$_{12}$ (left) and the possible AFM domains (right panels)~\cite{Troc2012}. $C_{3}^\pm \left(S_{6}^\pm \right)$ denotes the three-fold (improper) rotation. Each of domains has a toroidal moment, $\pm T \hat{z}$.}
\label{u3ru4al12} 
\end{figure}

The magnetic order of U$_3$Ru$_4$Al$_{12}$~\cite{Troc2012} is represented by the $A_{2u}$ and $E_{2u}$ irreducible representations. These representations are odd parity, and the former (latter) is polar (nonpolar). By representing one of the magnetic domains by the basis $\phi_{E_{2u}}+\phi_{A_{2u}} $, the other AFM domains are labeled as depicted in Fig~\ref{u3ru4al12}. The $\phi_{A_{2u}}$ mode corresponds to the toroidal moment $T_z$, and hence the electric current $j_z$ enables the switching between the AFM domains having different $\phi_{A_{2u}}$ components. On the other hand, the domains with the same $\phi_{A_{2u}}$ components cannot be switched by the electric current. Thus, the representation analysis is consistent with the Aizu species analysis in a partial toroidic case of U$_3$Ru$_4$Al$_{12}$. To control the AFM domains perfectly, we may use the magnetopiezoelectric effect, which is explained in Supplemental Material~\cite{supplemental}.

\textit{Read-out of AFM domains.} Following the symmetry analysis revealing an essential role of ferrotoroidic order for the electrical switching of AFM domains, we present a complete read-out of the domains using a functionality arising from the unidirectional nature of the ferrotoroidic order.

In an experiment of CuMnAs, an electrical read-out of AFM domains has been performed by measuring anisotropic magnetoresistance (AMR)~\cite{Wadley2016}. The AMR, however, cannot completely distinguish domains, because domains with opposite toroidal moments show the same AMR. On the other hand, we can make use of the \textit{unidirectional} property of the ferrotoroidic order to discern the AFM domains in a complete way.

The ferrotoroidic order induces a unidirectional anisotropy in various transport phenomena: In a nonlinear electric conductivity up to the second order $|\bm{E}|^2$ denoted by
    \begin{equation}
    j_\mu = \sigma_{\mu\nu} E_\nu + \tilde{\sigma}_{\mu\nu} E_\nu^2, \label{nonlinear_conductivity}
    \end{equation}
the ferrotoroidic moment $\bm{T} \parallel \hat{x}_\mu $ gives rise to a finite longitudinal component $\tilde{\sigma}_{\mu\mu}$, which changes sign between domains with opposite toroidal moment~\cite{Watanabe2018grouptheoretical}. Therefore, the nonlinear conductivity may distinguish the domain states of switchable antiferromagnets in a complete manner. Thus, both of the manipulation and detection of AFM states can be electrically carried out. 

The nonlinear conductivity $\tilde{\sigma}_{\mu\mu}$ indicates a dichromatic  transport. Indeed, dichromatic transport is an emergent physical property induced by the ferrotoroidic order. Although dichromatic transport has been observed in noncentrosymmetric systems under an external magnetic field~\cite{Rikken2001,PhysRevLett.94.016601,Ideue2017,Wakatsuki2017} and several ferromagnetic materials~\cite{Yasuda2016unidirectional,Olejnik2015unidirectional}, it may be realized in ferrotoroidic AFM states without a magnetic field. Such dichroism induced by the ferrotoroidic order is tunable by the current $\bm{j}$ through AFM domain switching. When we vary an electric current, a hysteretic behavior may be observed as a signal of AFM domain switching.

It is noteworthy that the manipulation and detection of the ferrotoroidic domain by a tunable electric current are realizable only in metallic systems in contrast to previous observations of ferrotoroidic order in magnetic insulators~\cite{VanAken2007a,Spaldin2008,Zimmermann2014a}. The toroidic domain in insulators can be manipulated by making use of the magnetoelectric effect. To be specific, the toroidic domain in insulators is inverted by simultaneously applying both electric and magnetic fields~\cite{VanAken2007a,Zimmermann2014a}. In contrast, the toroidic domains in metals are controllable by only injecting the electric current.

To summarize, we provide a general criterion of electrically switchable antiferromagnets based on the complementary use of the Aizu species and the representation theory. Both approaches unveil the direct correspondence between switchable AFM states and ferrotoroidic order. The concept of ferrotoroidic order uncovers functionalities of antiferromagnets and gives a clear viewpoint in AFM spintronics. It is desirable for further developments of AFM spintronics to explore the functionalities of various antiferrromagnets. On the basis of the symmetry analysis, we provided a list of electrically switchable antiferromagnets, which will be useful for future studies.

Recently, we became aware of an experiment of CuMnAs which demonstrated the switching and reading of the AFM domain states with opposite toroidal moments~\cite{Godinho2018electrically}. The domain states have been distinguished by the nonlinear Hall conductivity which is described by Eq.~\eqref{nonlinear_conductivity}. The toroidal moment $\bm{T} \parallel \hat{x}_\mu$ gives rise to a transverse nonlinear conductivity $\tilde{\sigma}_{\mu \nu}$~\cite{Gao2018}, and hence the experimental result~\cite{Godinho2018electrically} is consistent with our symmetry analysis.

\textit{Acknowledgments--- }The authors are grateful to M.~Kimata, S.~Nakatsuji, and S.~Suzuki for fruitful discussions. This work is supported by a Grant-in-Aid for Scientific Research on Innovative Areas ``J-Physics'' (Grant No.~JP15H05884) and ``Topological Materials Science'' (Grant No.~JP16H00991,~JP18H04225) from the Japan Society for the Promotion of Science (JSPS), and by JSPS KAKENHI (Grants No.~JP15K05164, No.~JP15H05745, and No.~JP18H01178). H.W. is supported by a JSPS research fellowship and supported by JSPS KAKENHI (Grant No.~18J23115).

\input{main_refs}
\clearpage

\renewcommand{\bibnumfmt}[1]{[S#1]}
\renewcommand{\citenumfont}[1]{S#1}
\renewcommand{\thesection}{S\arabic{section}}
\renewcommand{\theequation}{S\arabic{equation}}
\setcounter{equation}{0}
\renewcommand{\thefigure}{S\arabic{figure}}
\setcounter{figure}{0}
\renewcommand{\thetable}{S\arabic{table}}
\setcounter{table}{0}
\makeatletter
\c@secnumdepth = 2
\makeatother

\onecolumngrid

\onecolumngrid
\begin{center}
\vspace{1cm}
\textbf{\large Supplemental Material:\\ Symmetry analysis of electrical switching of antiferromagnets}
\end{center}

\section{Magnetic representation theory} \label{AppSection_magnetic_representation}
Here, we present a brief introduction to the representation theory for magnetic phase transitions~\cite{supplemental_Bertaut1968,supplemental_izyumov1991neutron} and apply the theory to the case of CuMnAs~\cite{supplemental_Wadley2013,supplemental_Wadley2016}. The magnetic representation, which describes symmetry of possible magnetic structures, is systematically obtained from a given crystal symmetry. A projection operator identifies the basis of the magnetic order in the way that the basis is symmetry-adapted to the irreducible representation.  

A magnetic order is accompanied by the loss of some symmetry operations of a given space group $\mathcal{G}$. Hence, the representation analysis based on the group theory is a powerful tool to investigate possible magnetic structures. The availability has been recognized in a lot of experimental works~\cite{supplemental_izyumov1991neutron}. Here we assume magnetic structures with no translational symmetry breaking ($\bm{Q}=\bm{0}$), where the magnetic unit cell is the same as the chemical cell. In this case, the magnetic structure is invariant to every translational operation of the space group $\mathcal{G}$, and thus the transformation property of the basis is determined by the point group $\bm{G}$ of the crystalline system. It is reasonable for the symmetry analysis of electrical switching of antiferromagnets to consider the point group symmetry, since applied external fields are uniform and cannot distinguish domain states induced by a translational symmetry breaking.

We denote by $\bm{m}^{(\alpha)}$ a magnetic moment localized at a crystallographic sublattice $\alpha$ in a unit cell. The magnetic basis $\{m_\nu^{(\alpha)}\}$ are transformed by the symmetry operation $g\in \bm{G}$ as 
    \begin{equation}
    g \left(m_\nu^{(\alpha)} \right) = \sum_{\beta \mu} m_\mu^{(\beta)}  [D^{(\bm{M})}(g)]_{\mu \nu} [ D^\textrm{(P)} (g)]_{\beta \alpha}, 
    \end{equation}  
where $D^{(\bm{M})}$ and $D^\textrm{(P)}$ are matrices which represent the transformation property of an axial vector and sublattice permutation, respectively. Therefore, the representation of the magnetic basis is written by the direct product,
    \begin{equation}
    \Gamma_{\bm{G}}^\mathrm{mag}\left(\bm{H} \right) =  \Gamma_{\bm{G}}^\mathrm{P} \left(\bm{H} \right)\otimes  \Gamma_{\bm{G}}^{\bm{M}}.\label{magnetic_representation_appendix}
    \end{equation}
$\Gamma_{\bm{G}}^\mathrm{P}$ represents the sublattice permutation representation  with the site-symmetry group $\bm{H}$. $\Gamma_{\bm{G}}^{\bm{M}}$ denotes the axial vector representation. All the magnetic structures constructed from $\{m_\nu^{(\alpha)}\}$ break the time-reversal symmetry, since the representation $\Gamma_{\bm{G}}^{\bm{M}} \left( \Gamma_{\bm{G}}^\textrm{P} \right)$ shows the odd (even) parity under the time-reversal operation.

The symmetry-adapted basis of the magnetic order are given by irreducible representations $\{ \Gamma_{\bm{G}}^{(a)} \}$. Now, we decompose the magnetic representation~\eqref{magnetic_representation_appendix} to identify which irreducible representation is comprised. The decomposition is given by 
    \begin{equation}
    \Gamma_{\bm{G}}^\mathrm{mag} \left(\bm{H} \right)= \sum_{a} q_a \Gamma_{\bm{G}}^{(a)}.
    \end{equation}
The independent magnetic basis of the representation $\Gamma_{\bm{G}}^{(a)}$ is as many as $q_a$. The coefficient $q_a$ is given by
    \begin{equation}
      q_a = \frac{1}{|\bm{G}|} \sum_{g\in \bm{G}} \chi_a^\ast \left(g \right) \chi_\textrm{mag} \left(g \right), \label{frequency_calculation_appendix}
    \end{equation}
where $|\bm{G}|$ is the order of the point group $\bm{G}$. $\chi_a \left(g \right)$ and $\chi_\mathrm{mag} \left(g \right)$ are characters of the representations $\Gamma_{\bm{G}}^{(a)}$ and $ \Gamma_{\bm{G}}^\mathrm{mag}$, respectively. The character $\chi_\mathrm{mag} \left(g \right)$ is obtained by multiplying the character of the representation $\Gamma_{\bm{G}}^\textrm{P}$ and that of $ \Gamma_{\bm{G}}^{\bm{M}}$ owing to Eq.~\eqref{magnetic_representation_appendix}.

 The magnetic basis of the irreducible representation $\Gamma_{\bm{G}}^{(a)}$, labeled by $\xi$, is given by the projection operator
    \begin{equation}
      \hat{P}_\xi^{(a)} = \frac{\mathrm{dim} \Gamma_{\bm{G}}^{(a)}}{|\bm{G}|}\sum_{g\in \bm{G}} [ D^{(a)} (g)]_{\xi\xi}^\ast ~g,\label{projection_operator_appendix}
    \end{equation}
where $\xi=1,2,\cdots,\mathrm{dim} \Gamma_{\bm{G}}^{(a)}$. In particular, for an one-dimensional irreducible representation, the projection operator is simplified as
    \begin{equation}
      \hat{P}^{(a)} = \frac{1}{|\bm{G}|}\sum_{g\in \bm{G}} \chi_a^\ast \left(g \right) ~g,\label{projection_operator2_appendix}
    \end{equation}
where $\chi_a \left( g\right) = \mathrm{Tr}D^{(a)} (g)$. Thus, we can complete all the possible magnetic structures by the representation theory technique.

\subsection{Application to CuMnAs} \label{AppSection_CuMnAs}
We apply the representation theory to CuMnAs, where the antiferromagnetic (AFM) domain switching has been demonstrated~\cite{supplemental_Wadley2013}. The compound shows a ferrotoroidic order in the AFM phase. Here, we investigate possible magnetic structures of CuMnAs and clarify the relation between the toroidal moment and the magnetic mode for the realized AFM order.

The magnetic sites, Mn atoms, are positioned at the crystallographic position with the site-symmetry group $\bm{H} = 4mm$, while the crystal group is $\bm{G} =4/mmm$. The coset decomposition of $\bm{G} $ by $\bm{H} $ is obtained as,
  \begin{equation}
      \bm{G} = I \bm{H} +P \bm{H}, \label{app_CuMnAs_decomposition}
    \end{equation}  
where $I$ and $P$ represent the identity operation and parity operation, respectively. The number of the sublattice is the ratio $|\bm{G}|/|\bm{H}|=2$ obtained from the coset decomposition~\eqref{app_CuMnAs_decomposition}.

Now, we examine a transformation property of the sublattice permutation. The matrix element of $D^\mathrm{(P)} (g)$ is given by
    \begin{equation}
      [ D^\mathrm{(P)} (g) ]_{a b} = \delta \left( a, g(b)\right), \label{permutation_matrix_appendix}
    \end{equation}
and $\delta \left( a, b\right)$ is defined as 
    \begin{equation}
      \delta \left( a, b\right)=
      \begin{cases}
      1&\text{for $a=b$},\\
      0&\text{for $a\neq b$},
      \end{cases}
    \end{equation}
which is parametrized by the sublattice indexes $a$ and $b$. In the case of CuMnAs, the two sublattices are interchanged by the parity operation $g=P$. Thus, the representation matrix is given by 
\begin{equation}
  D^\mathrm{(P)} (P)=
  \begin{pmatrix}
  0&1\\
  1&0
  \end{pmatrix},\label{permutation_matrix_example_parity}
\end{equation}
and therefore the character is $\chi_\mathrm{P} \left(P \right)= \textrm{Tr}~D^{(P)} \left( P \right) =0$. It follows that any sublattice does not return to its crystallographic position by the operation $P$. The condition $\chi_\mathrm{P} \left(P \right)=0$ represents the locally noncentrosymmetric property of Mn atoms in CuMnAs. Similarly, characters of the symmetry operation $g\in \bm{G}$ are obtained as in Table~\ref{character_for_CuMnAs_magnetic_representation}.

The characters of the axial vector representation $\chi_{\bm{M}} \left(g \right)$ are obtained by the trace of the representation matrices of an axial vector. In the case of the group $\bm{G}=4/mmm$, the character $\chi_{\bm{M}} \left(g \right)$ is given by a summation of the characters of the $A_{2g}$ and $E_{g}$ irreducible representations. This is because the axial vector representation $\Gamma_{\bm{G}}^{\bm{M}}$ is given by the direct sum of $A_{2g}$ and $E_g$. In fact, magnetization $M_z$ and $\{ M_x, M_y\} $ belong to the representations $A_{2g}$ and $E_{g}$, respectively. 

    \begin{table}[htbp]
      \caption{The character table in the crystal group $\bm{G}=4/mmm$ for the representation $\Gamma_{\bm{G}}^\mathrm{P}$, $\Gamma_{\bm{G}}^{\bm{M}}$, and $\Gamma_{\bm{G}}^\mathrm{mag}$. The magnetic sites are supposed to be Mn atoms of CuMnAs. The conventional notation is adopted for the symmetry operations of $\bm{G}=4/mmm$~\cite{supplemental_inui1990group}.}
    \label{character_for_CuMnAs_magnetic_representation}
    \centering 
    \begin{tabular}{lcccccccccc}
      \hline \hline
      $g \in \bm{G}$&$E  $&$  2C_{4}  $&$ C_{2z}  $&$  2C'_{2}  $&$ 2C'_{2}  $&$P  $&$  2S_{4}  $&$ \sigma_{h}  $&$  2\sigma_{v}  $&$ 2\sigma_{d}  $ \\ \hline
      $\chi_\mathrm{P} \left(g \right)$&2&2&2&0&0&0&0&0&2&2 \\
      $\chi_{\bm{M}} \left(g \right)$&3&1&-1&-1&-1&3&1&-1&-1&-1 \\
      $\chi_\mathrm{mag} \left(g \right)$&6&2&-2&0&0&0&0&0&-2&-2 \\
      \hline\hline
    \end{tabular}
    \end{table}

The coefficients $\{ q_a\}$ calculated by Eq.~\eqref{frequency_calculation_appendix} give the decomposition of the magnetic representation as
    \begin{equation}
      \Gamma_{\bm{G}}^\mathrm{mag} \left(\bm{H} \right) = A_{2g} +A_{1u} +E_g +E_u, \label{CuMnAs_irreducible_decompose_by_characteres}
    \end{equation}
among which the $E_u$ irreducible representation corresponds to the AFM order of CuMnAs~\cite{supplemental_Wadley2013,supplemental_Wadley2016}.

The symmetry-adapted basis are obtained by the projection operator~\eqref{projection_operator_appendix}. Taking the basis of the $E_u$ representation as $\{ x,y\}$, we accordingly obtain matrix elements of the representation matrix $D^{(E_u)}$ in Eq.~\eqref{projection_operator_appendix}. The projection operator for the $E_u \left(x \right)$ basis identifies the magnetic structure
    \begin{equation}
    \bm{m}^{(1)}= m \hat{y},~ \bm{m}^{(2)}= -m \hat{y}, \label{toroidic_mode_along_x_appendix}
    \end{equation}
where $\bm{m}^{(1)}$ and $\bm{m}^{(2)}$ represent magnetic moments localized at the two sublattices. The toroidal moment $T_0  \hat{x}$ also belongs to the $E_u \left(x \right)$ basis, and hence the magnetic structure~\eqref{toroidic_mode_along_x_appendix} is induced by the electric current $\bm{j}\parallel \hat{x}$. Similarly, we obtain the the magnetic structure belonging to the $E_u \left(y \right)$ basis as
    \begin{equation}
    \bm{m}^{(1)}= -m \hat{x},~ \bm{m}^{(2)}= m \hat{x}, \label{toroidic_mode_along_y_appendix}
    \end{equation}
which corresponds to the toroidal moment $T_0  \hat{y}$. The result of the representation analysis is consistent with the microscopic study for the AFM Edelstein effect~\cite{supplemental_Zelezny2017}.

Note that we may obtain the irreducible decomposition of $\Gamma_{\bm{G}}^\mathrm{P}$ without making use of the permutation matrices such as Eq.~\eqref{permutation_matrix_example_parity}. The definition~\eqref{permutation_matrix_appendix} says that the permutation representation is the induced representation of the identity representation $\Gamma_{\bm{H}}^{(1)} $ on the group $\bm{G}$,
    \begin{equation}
      \Gamma_{\bm{G}}^\mathrm{P} \left(\bm{H} \right)= \Gamma_{\bm{H}}^{(1)} \uparrow{\bm{G}} =\sum_{a} p_a \Gamma_{\bm{G}}^{(a)}, \label{permutation_representation_decomposition_appendix}
    \end{equation}
In general, irreducible representations $ \{ \Gamma_{\bm{G}}^{(a)} \}$ are reducible in the subgroup $\bm{H} \left( \subset \bm{G} \right)$, and hence those representations are decomposed by the irreducible representations of $\bm{H}$ as
    \begin{equation}
    \Gamma_{\bm{G}}^{(a)} \downarrow \bm{H}= \sum_{b} {p'_b}^{(a)} \Gamma_{\bm{H}}^{(b)}.
    \end{equation}
Following the Frobenius reciprocity~\cite{supplemental_inui1990group}, a useful formula is obtained as
    \begin{equation}
    p_a = {p'_1}^{(a)},
    \end{equation}  
where the right-hand-side is obtained by the compatibility relation between the groups $\bm{G}$ and $\bm{H}$. In the case of CuMnAs $\left(\bm{G}=4/mmm, \bm{H}=4mm \right)$, the compatibility of the irreducible representation is shown in Table~\ref{compatibility_D4h_and_C4v_appendix}. Therefore, the coefficients in Eq.~\eqref{permutation_representation_decomposition_appendix} are obtained as
    \begin{equation}
      p_{a}=
      \begin{cases}
      1&\text{for $a= A_{1g}, A_{2u}$},\\
      0&\text{otherwise}.
      \end{cases}
    \end{equation}
Thus, $\Gamma_{\bm{G}}^\mathrm{P}$ is given by
    \begin{equation}
    \Gamma_{\bm{G}}^\mathrm{P}= A_{1g}+A_{2u}.
    \end{equation}
As demonstrated above, the irreducible decomposition of $\Gamma_{\bm{G}}^\mathrm{P}$ is determined by only the compatibility relation between $\bm{G}$ and $\bm{H}$. Accordingly, with the use of product rules for the irreducible representations we obtain $\Gamma_{\bm{G}}^\mathrm{mag}$ as
  \begin{align}
  \Gamma_{\bm{G}}^\textrm{mag} (\bm{H})
    &= \Gamma_{\bm{G}}^\mathrm{P} (\bm{H}) \otimes \Gamma_{\bm{G}}^{\bm{M}},\\
    &= \left( A_{1g}+A_{2u} \right) \otimes \left( A_{2g}+E_{g} \right),\\
    &= A_{2g}+E_{g}+A_{1u}+E_u,
  \end{align}
which is the same result as Eq.~\eqref{CuMnAs_irreducible_decompose_by_characteres}. 

    \begin{table}[htbp]
    \caption{The compatibility relation of the irreducible representations of $\bm{G}= 4/mmm$ in the case of the symmetry reduction $\bm{G}= 4/mmm \rightarrow \bm{H}= 4mm$. The $A_{1}$ irreducible representation is the identity representation of the group $4mm$.}
    \label{compatibility_D4h_and_C4v_appendix}
    \centering 
      \begin{tabular}{ccccccccccc}
      \hline\hline 
      $\Gamma_{\bm{G}}$
      &$A_{1g}$&$A_{2g}$&$B_{1g}$&$B_{2g}$&$E_{g}$
      &$A_{1u}$&$A_{2u}$&$B_{1u}$&$B_{2u}$&$E_{u}$ \\ 
      $\Gamma_{\bm{G}} \downarrow \bm{H}$
      &$A_{1}$&$A_{2}$&$B_{1}$&$B_{2}$&$E_{}$
      &$A_{2}$&$A_{1}$&$B_{2}$&$B_{1}$&$E_{}$ \\ 
      \hline\hline
    \end{tabular}
    \end{table}

\section{Extension to noncentrosymmetric systems} \label{AppSection_extension_to_noncentrosymmetric}
Our symmetry analysis can be straightforwardly extended to ferromagnetic or AFM order in noncentrosymmetric systems, while in the main text we focus on the AFM order in centrosymmetric crystalline systems. In this section, we introduce a symmetry analysis based on the Aizu species and the representation theory in noncentrosymmetric systems, and apply the extended scheme to strained (Ga,Mn)As and MnSiN$_2$ as examples of switchable magnets with noncentrosymmetric crystalline structures. 

The Aizu species analysis is extended in a straightforward way. When the system undergoes the magnetic phase transition which is switchable by the electric current, its species should be full or partial toroidic. Thus, the presence of the toroidal moment is a criterion of the electrical switching, irrespective of whether the crystalline structure is centrosymmetric or noncentrosymmetric. An important difference of noncentrosymmetric systems from centrosymmetric systems is the following. The species of switchable magnets can be full-magnetic, that is, magnetic structures can be ferromagnetic, since the $\mathcal{PT}$ symmetry is absent in the noncentrosymmetric systems. A switchable domain state may comprise ferromagnetic moment in addition to the toroidal moment.

As for the representation analysis of noncentrosymmetric systems, we obtain possible magnetic basis in the same manner as in Appendix~\ref{AppSection_magnetic_representation}. By supposing magnetic sites with site-symmetry group $\bm{H}$, magnetic representation of crystal group $\bm{G}$ is obtained as Eq.~\eqref{magnetic_representation_appendix}. The parity of each magnetic basis cannot be determined anymore owing to the fact that $\bm{G}$ is noncentrosymmetric. Below we discuss ferromagnet and antiferromagnet on the basis of the representation theory.

First, we present the criterion for the current-induced switching of ferromagnets. In contrast to centrosymmetric systems, the (ferromagnetic) Edelstein effect is allowed in noncentrosymmetric crystals~\cite{supplemental_Garate2009,supplemental_Ciccarelli2015}. The formula for the Edelstein effect is written as
    \begin{equation}
    M_\mu = \kappa_{\mu\nu}^\mathrm{FM} j_\nu,
    \end{equation}
where the susceptibility tensor has no sublattice degree of freedom in contrast to the AFM Edelstein effect [Eq.~(3) in the main text]. The magnetic representation of the ferromagnetic order is same as the axial vector representation,
    \begin{equation}
    \Gamma_{\bm{G}}^\mathrm{FM} = \Gamma_{\bm{G}}^{\bm{M}}  \subset \Gamma_{\bm{G}}^\mathrm{mag} \left(\bm{H} \right),
    \end{equation}
which is obtained by replacing the permutation representation $\Gamma_{\bm{G}}^\mathrm{P}$ in Eq.~\eqref{magnetic_representation_appendix} with the identity representation $\Gamma_{\bm{G}}^{(1)}$.

Here we impose the symmetry of a paramagnetic state on the susceptibility $\kappa_{\mu\nu }^\mathrm{FM}$. The symmetry operations $g\in \bm{G}$ give constraints to the susceptibility tensor $\kappa_{\mu\nu}^\mathrm{FM}$ in accordance with Neumann's principle. The representation of the susceptibility $\kappa_{\mu\nu}^\mathrm{FM}$ is decomposed as
      \begin{equation}
       \Gamma_{\bm{G}}^\mathrm{FM} \otimes  \Gamma_{\bm{G}}^{\bm{j}}= \sum_\alpha p_\alpha  \Gamma_{\bm{G}}^{(\alpha)}, \label{FM_Edelstein_irreducible_decomposition}
      \end{equation} 
where $p_1$ for the identity representation $\Gamma_{\bm{G}}^{(1)}$ gives the number of independent coefficients of the tensor $\kappa_{\mu\nu}^\mathrm{FM}$. The condition $p_1 \neq 0 $ can be represented by 
      \begin{equation}
      \sum_{g\in \bm{G}} \chi_{\bm{j}}^\ast \left(g \right) \chi_\mathrm{FM} \left(g \right) \neq 0,
      \end{equation}
due to Eq.~\eqref{frequency_calculation_appendix}. $\chi_{\bm{j}} \left(g \right)$ and $\chi_\mathrm{FM} \left(g \right)$ are the character of the representations $\Gamma_{\bm{G}}^{\bm{j}}$ and $\Gamma_{\bm{G}}^\mathrm{FM}$, respectively.

The ferromagnetic representation should share the same irreducible representation with the polar vector representation $\Gamma_{\bm{G}}^{\bm{j}}$ when the ferromagnetic order is induced by the electric current $\bm{j}$. In other words, the ferromagnetic representation should comprise the toroidal moment representation which is the same as $\Gamma_{\bm{G}}^{\bm{j}}$. Thus, the toroidal moment is necessary for the current-induced ferromagnetic domain switching,  although the attention was not paid in the spintronics studies. The crystal groups, where the Edelstein effect is allowed, are called as gyrotropic~\cite{supplemental_Garate2009}.

Next, we show the criterion for the switching of antiferromagnets. The representation of AFM order is obtained by subtracting the ferromagnetic representation $\Gamma_{\bm{G}}^\mathrm{FM}$ from the magnetic representation $\Gamma_{\bm{G}}^\mathrm{mag}$ owing to the relation,
    \begin{equation}
    \Gamma_{\bm{G}}^\mathrm{mag}\left(\bm{H} \right)= \Gamma_{\bm{G}}^\mathrm{FM}+\Gamma_{\bm{G}}^\mathrm{AFM}\left(\bm{H} \right). \label{AFM_representation_appendix}
    \end{equation}
Similarly, the representation of the AFM Edelstein susceptibility is decomposed as 
    \begin{equation}
     \Gamma_{\bm{G}}^\mathrm{AFM} \left(\bm{H} \right)\otimes  \Gamma_{\bm{G}}^{\bm{j}}= \sum_\alpha r_\alpha  \Gamma_{\bm{G}}^{(\alpha)}.
    \end{equation} 
The AFM Edelstein effect is allowed when satisfying $r_1 \neq 0$. The condition $r_1 \neq 0$ can be recast   
    \begin{equation}
    \sum_{g\in \bm{G}} \chi_{\bm{j}}^\ast \left(g \right) \left[  \chi_\mathrm{mag} \left(g \right) -\chi_\mathrm{FM} \left(g \right) \right]  \neq 0,
    \end{equation}
which is derived from Eq.~\eqref{AFM_representation_appendix}. In the case of $r_1\neq 0 $, the AFM representation comprises the toroidal moment representation, that is, $\Gamma_{\bm{G}}^{\bm{j}}$. The toroidal moment which a realized AFM state comprises is inverted to be parallel to the injected electric current. Thus, the criterion is the same as that for the switching of ferromagnets.

In the following, we discuss two examples: stained (Ga,Mn)As is a noncentrosymmetric ferromagnet whereas MnSiN$_2$ is a noncentrosymmetric antiferromagnet.

\subsection{Strained \textrm{(Ga,Mn)As}} \label{AppSection_strainedGaMnAs}
(Ga,Mn)As is a ferromagnetic semiconductor, crystallizing in the zinc-blende structure~\cite{supplemental_Ohno1998}. Let us assume that the magnetic atoms (Mn) are positioned at Ga sites. The magnetic sites have no sublattice degree of freedom, and the magnetic representation is obtained as 
    \begin{equation}
    \Gamma_{\bm{G}}^\mathrm{mag}\left(\bm{H} \right) =\Gamma_{\bm{G}}^\mathrm{FM}=\Gamma_{\bm{G}}^{\bm{M}}. \label{GaMnAs_magnetic_rep_appendix}
    \end{equation}
The Edelstein effect is not allowed in (Ga,Mn)As, since the crystal group of the zinc-blende structure is $\bm{G}=\bar{4}3m$ which is noncentrosymmetric but non-gyrotropic. In fact, the magnetic representation~\eqref{GaMnAs_magnetic_rep_appendix} given by $ \Gamma_{\bm{G}}^\mathrm{mag} = T_1$ differs from the polar vector representation $ \Gamma_{\bm{G}}^{\bm{j}} =T_2$. Thus, the criterion for the switching magnetic domain is not satisfied.

Now, we suppose that the crystal structure is deformed from cubic to tetragonal by applying strain represented by $\epsilon_{zz}$. Accordingly, the crystal group is transformed into $\bm{G}'=\bar{4}2m$. The applied stain also reduces the representations of axial and polar vectors as
    \begin{align}
    \begin{aligned}
    &\Gamma_{\bm{G}}^{\bm{M}} \downarrow 
    \bm{G}' =A_{2}+E,\\
    &\Gamma_{\bm{G}}^{\bm{j}} \downarrow \bm{G}' =B_{2}+E,
    \end{aligned}
    \end{align}
where we use the compatibility relation between $\bm{G}$ and $\bm{G}'$. The representations $\Gamma_{\bm{G}}^\mathrm{mag}$$\left(  =\Gamma_{\bm{G}}^{\bm{M}}  \right)$ and $\Gamma_{\bm{G}}^{\bm{j}}$ comprise the same representation $E$. Therefore, $p_1=1$ in Eq.~\eqref{FM_Edelstein_irreducible_decomposition}, since the identity representation $\Gamma_{\bm{G}'}^{(1)}=A_1$ is obtained from the product representation $E \otimes E$. Indeed, the product is decomposed as
      \begin{equation}
      E \otimes E =A_1+A_2+B_1+B_2.
      \end{equation}
Thus, the Edelstein effect is allowed in the strained system. In other words, the crystal group is changed from non-gyrotropic into gyrotropic by applying the strain. 

By using the projection operator associated with the $E$ irreducible representation, the ferromagnetic moment $M_x$ and $M_y$ are identified to be symmetry-adapted to the electric current $j_y$ and $j_x$, respectively. When the ferromagnetic moment $M_x >0$ is induced by the electric current $j_y>0$, $M_y>0$ has to be induced by $j_x>0$ because of the fourfold improper rotations of $\bm{G}'$. The elements of the tensor $\kappa_{\mu\nu}^\mathrm{FM}$ are given by 
    \begin{equation}
      \begin{cases}
      \kappa_{xy}^\mathrm{FM}= \kappa_{yx}^\mathrm{FM} \neq 0,&\\
      \kappa_{\mu\nu}^\mathrm{FM}=  0 &\left( \mu, \nu \right) \text{ for otherwise}.\\
      \end{cases}\label{strained_GaMnAs_Edelstein_appendix}
    \end{equation}
This form of the susceptibility corresponds to the Dresselhaus-type spin-momentum coupling~\cite{supplemental_Chernyshov2009}.

We also perform symmetry analysis based on the Aizu species. When the ferromagnetic moment is aligned along the $x$ axis in the strain-free (Ga,Mn)As, the species and its property are obtained as
      \begin{center}
          \begin{tabular}{ccccc} \hline \hline 
          \multirow{2}{*}{$\aizu{\bar{4}3m1'}{\bar{4}2'm'}$}&$\hat{\epsilon}$&$\bm{P}$&$\bm{M}$&$\bm{T}$ \\  
          &P&Z&F&Z\\ \hline\hline
          \end{tabular}
      \end{center}
The species is full magnetic and zero toroidic. It turns out that the ferromagnetic moment of the unstrained (Ga,Mn)As is not switchable by the electric current, while it can be inverted by the magnetic field.

On the other hand, the species of a strained system with in-plane ferromagnetic moment is described by
      \begin{center}
          \begin{tabular}{ccccc} \hline \hline 
          \multirow{2}{*}{$\aizu{\bar{4}2m1'}{2'2'2 \Braket{x}}$}&$\hat{\epsilon}$&$\bm{P}$&$\bm{M}$&$\bm{T}$ \\  
          &P&Z&F&F\\ \hline\hline
          \end{tabular}
      \end{center}
where $\Braket{x}$ indicates that the twofold rotation axis is the $x$ or $y$ axis. The species of the strained (Ga,Mn)As turns into full toroidic species. Thus, the in-plane ferromagnetic order is perfectly controllable not only by the magnetic field but also by the electric current. The Aizu species analysis is consistent with the representation analysis.

\subsection{MnSiN\texorpdfstring{$_2$}{TEXT}} \label{AppSection_MnSiN2}
MnSiN$_2$ crystallizes in an orthorhombic structure with point group $\bm{G}=mm2$ (space group: P$na2_1$, No.~33)~\cite{supplemental_Esmaeilzadeh2006}. Magnetic sites (Mn) are located at the crystallographic position whose site-symmetry group is $\bm{H} =1$. The compound may be a candidate for antiferromagnetic spintronics devices~\cite{supplemental_Baltz2018}, since it undergoes AFM phase transition with $\bm{Q}=\bm{0}$ below a high N\'eel temperature $T_\mathrm{N} \sim \mr{500}{K}$~\cite{supplemental_Esmaeilzadeh2006}.

We approximate the magnetic structure as a collinear AFM order parallel to the $z$ axis for simplicity, although magnetic moments are almost aligned along the $z$ axis with small canting~\cite{supplemental_Esmaeilzadeh2006}. This simplification does not change the conclusion. The magnetic representation is obtained as 
  \begin{equation}
  \Gamma_{\bm{G}}^\textrm{mag} (\bm{H}) 
    =3A_{1}+3A_{2}+3B_{1}+3B_{2}. \label{MnSiN2_magnetic_representation}
  \end{equation}
The ferromagnetic representation is given by 
    \begin{equation}
    \Gamma_{\bm{G}}^\textrm{FM} 
      =A_{2}+B_{1}+B_{2}. \label{MnSiN2_ferromagnetic_representation}
    \end{equation}
Therefore, the AFM representation is
   \begin{equation}
    \Gamma_{\bm{G}}^\textrm{AFM} \left(\bm{H} \right)
      =3A_{1}+2A_2+2B_{1}+2B_{2}, \label{MnSiN2_antiferromagnetic_representation}
    \end{equation}
due to Eq.~\eqref{AFM_representation_appendix}. The polar representation $\Gamma_{\bm{G}}^{\bm{j}} =A_1+B_1+B_2$ is comprised in Eqs.~\eqref{MnSiN2_ferromagnetic_representation} and \eqref{MnSiN2_antiferromagnetic_representation}. Thus, both of the FM and AFM Edelstein effects are allowed.

The magnetic order reported in the experiment~\cite{supplemental_Esmaeilzadeh2006} is represented by one of the $B_1$ modes of the AFM representation~\eqref{MnSiN2_antiferromagnetic_representation}. The AFM moment of MnSiN$_2$ can be inverted by an electric current $j_x$, since the basis of the $B_1$ irreducible representation can be taken as a toroidal moment $T_x$.

The feasibility of the electrical switching is also supported by the Aizu species analysis. The species of MnSiN$_2$ is given by
    \begin{center}
        \begin{tabular}{ccccc} \hline \hline 
        \multirow{2}{*}{$\aizu{mm21'}{m'm2'}$}&$\hat{\epsilon}$&$\bm{P}$&$\bm{M}$&$\bm{T}$ \\  
        &Z&Z&F&F\\ \hline\hline
        \end{tabular}
    \end{center}
The species is full toroidic with the toroidal moment $\bm{T}\parallel \hat{x}$, and hence the electric current $j_x$ is coupled to the toroidal moment and switches the AFM domains.

The species is also full magnetic, which indicates that the AFM state of MnSiN$_2$ can hold a net magnetization and the net magnetization of AFM domains are different from each other. The allowed net magnetization is along the $y$ axis due to the preserved mirror symmetry for the $zx$plane.

According to the Aizu species of MnSiN$_2$, magnetization $M_y$ induced by a magnetic field may invert the AFM moment, since the induced magnetization $M_y$ can be coupled to the toroidal moment $T_x$ arising from the AFM moment. Such an indirect switching of the AFM order with the magnetic field is forbidden in centrosymmetric crystals because of the $\mathcal{PT}$ symmetry.

\section{Notes for the Aizu species analysis} \label{app_notification_on_Aizu_species}

Given various order parameters such as electric polarization and ferromagnetic moment, Aizu species, an ensemble of pairs of disorder phase and order phase, is classified in a corresponding way. In this section, we note the convention in Ref.~\cite{supplemental_Litvin2008} for the Aizu species classification by time-reversal-even ($\mathcal{T}$ even) physical quantities, \textit{e.g.}~electric polarization $\bm{P}$ and strain $\hat{\epsilon}$.

An order parameter of the $\mathcal{T}$ even order is equivalent between the domain states connected by the time-reversal operation. This twofold degeneracy has been neglected in the Aizu species classification of Ref.~\cite{supplemental_Litvin2008}. For instance, an Aizu species written as    
      \begin{equation}
      \aizu{mmm1'}{m'm2'},\label{aizu_species_full_electric}  
      \end{equation} 
is characterized as ``full (F)'' electric in Ref.~\cite{supplemental_Litvin2008}. The domain states therefore seem to be perfectly controlled by an electric field $\bm{E}$, since the field $\bm{E}$ is conjugated to the electric polarization $\bm{P}$. On the other hand, using the coset decomposition of the group $mmm1'$ by $m'm2'$, we obtain electric polarizations in each domain $\{s_i\}$ as
      \begin{equation}
      (s_i, \bm{P}_i)= (s_1, \bm{P}_0),~(s_2, \bm{P}_0),~(s_3, -\bm{P}_0),~(s_4, -\bm{P}_0),
      \end{equation}
where we suppose $\bm{P}_0 \parallel \hat{z}$ without loss of generality. The domains are related with each other as
      \begin{equation}
      s_2=\theta s_1,~ s_3=C_{2x} s_1,~ s_4= \theta C_{2x} s_1.
      \end{equation}
The domains connected by the time-reversal operation $\theta$ cannot be distinguished by the electric polarization $\bm{P}$. Thus, the species should be characterized as partial electric in the rigorous sense. Especially, it is important for our symmetry analysis of the switchable antiferromagnets to distinguish the domains connected by the operation $\theta$, since those domains may be discerned by the toroidal moment and inverted by applying the electric current.

\section{Partial toroidic property of magnetic honeycomb lattice} \label{app_magnetostrictive}

\begin{figure*}[htbp] 
\centering 
\includegraphics[width=160mm,clip]{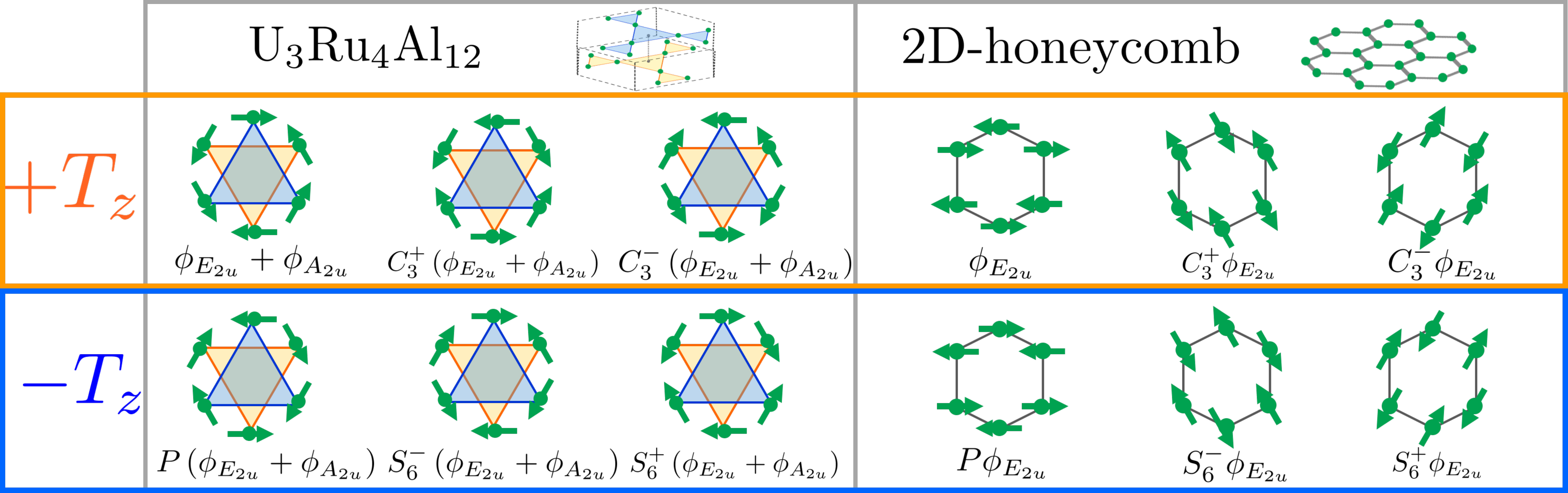}
\caption{The magnetic domains of U$_3$Ru$_4$Al$_{12}$~\cite{supplemental_Troc2012} (left panels) and two-dimensional honeycomb lattice (right panels). $C_{3}^\pm \left(S_{6}^\pm \right)$ denotes the threefold rotation (rotatory inversion) operation. U$_3$Ru$_4$Al$_{12}$ has a toroidal moment, $\pm T \hat{z}$, while it vanishes in the honeycomb lattice without taking the magnetostrictive effect into account.}
\label{combined} 
\end{figure*}

In this section, we consider a fictitious example of a partial toroidic species: a collinear AFM order of a honeycomb lattice. By comparing the honeycomb lattice to U$_3$Ru$_4$Al$_{12}$, which also belongs to a partial toroidic species, we illuminate a complementary role of the representation analysis and the Aizu species analysis.

Let us assume that the honeycomb lattice (the crystal group $\bm{G} =6/mmm$), hosting two sublattices with the site-symmetry group $\bm{H}=\bar{6}2m$, undergoes a collinear AFM order aligned along the twofold axis in the $xy$ plane as shown in the right panels of Fig~\ref{combined}. Such magnetic order may realize in magnetic honeycomb systems such as transition-metal trichalcogenides~\cite{supplemental_Chittari2016,supplemental_Gong2017}.

The assumed AFM order belongs to the same Aizu species as that of  U$_3$Ru$_4$Al$_{12}$ [Eq.~(14) in the main text]. Therefore, we may expect that the domains are partially-controllable by the electric current $j_z$ as in the case of U$_3$Ru$_4$Al$_{12}$. The magnetic representation theory of the honeycomb lattice, however, leads to
      \begin{equation}
      \Gamma_{\bm{G}}^\textrm{mag} \left(\bm{H} \right)
      =A_{2g}+E_{1g}+B_{2u}+E_{2u}, \label{honeycomb_magnetic_representation}
      \end{equation}
which do not comprise the polar representation $\Gamma_{\bm{G}}^{\bm{j}} =A_{2u}+E_{1u}$ in contrast to the magnetic representation of U$_3$Ru$_4$Al$_{12}$ [Eq.~(15) in the main text]. As shown in Fig.~\ref{combined}, the AFM structure of the honeycomb lattice actually contains only a nonpolar mode obtained as the $E_{2u}$ mode, while the AFM order of U$_3$Ru$_4$Al$_{12}$ is represented by a polar mode, that is, the $A_{2u}$ mode in addition to the $E_{2u}$ mode. Then, the current $j_z$ is not linearly coupled to the AFM order. Thus, the AFM domains of the honeycomb lattice are not switchable by the AFM Edelstein effect.

The criterion for the controllability, that is, whether the AFM order comprises a toroidic moment depends on the site-symmetry of magnetic sites in a partial toroidic species. The representation theory and the Aizu species analysis are complementary: The presence or absence of a toroidal moment in an AFM state should be checked by the representation analysis, while the controllability of domains is understood by the Aizu species analysis.

To keep the consistency between the Aizu species analysis and the representation theory analysis as for the honeycomb lattice, we need to consider a magnetostrictive effect. In Eq.~\eqref{honeycomb_magnetic_representation}, we assume that the magnetic representation is determined by the crystalline structure in the paramagnetic phase. A magnetic phase transition, however, may give rise to a structural change through a magnetic-elastic coupling, namely, magnetostrictive effect. The structural change transforms the magnetic representation into that in a lowered crystal symmetry.

Now, we consider the AFM phase which has the polar axis along the $z$ axis and the mirror symmetry for the $zx$ plane. Owing to the magnetostrictive effect, the strain $\epsilon_{xx}-\epsilon_{yy}$ is induced by the AFM order. This is phenomenologically understood by the Landau's free energy,
        \begin{equation}
          \mathcal{F} = a \eta_\mathrm{AF}^2 +c \eta_\mathrm{AF}^4 + \lambda \eta_\mathrm{AF}^2 \eta_{\epsilon} +\cdots,   \label{free_energy_expansion_appendix}
        \end{equation}
where $\eta_\mathrm{AF}$ and $\eta_{\epsilon}$ represent the AFM order parameter and the induced strain. $\lambda$ represents the magneto-elastic coupling. When the AFM phase transition occurs with a non-negligible coupling $\lambda$, the crystal structure is deformed from hexagonal to orthorhombic as $\bm{G}=6/mmm$ to $\bm{G}'=mmm$. Similarly, the site-symmetry group of the sublattice is transformed as $\bm{H}=\bar{6}2m$ to $\bm{H}'=2mm$. The magnetic representation of the honeycomb lattice is reduced from Eq.~\eqref{honeycomb_magnetic_representation} to
      \begin{equation}
        \Gamma_{\bm{G}'}^\textrm{mag} (\bm{H}') 
        =B_{1g} +\left(  B_{2g}+B_{3g} \right)+B_{2u} +\left(  A_{u}+B_{1u} \right). \label{honeycomb_magnetic_representation_corrected}
      \end{equation}
The $E_{2u}$ irreducible representation becomes reducible in the descended group $\bm{G}'$, and the irreducible decomposition is given by
    \begin{equation}
    E_{2u} \downarrow{\bm{G}'}= A_{u}+B_{1u}.
    \end{equation}
The toroidal moment $\bm{T}\parallel \hat{z}$ belongs to the $B_{1u}$ irreducible representation of $\bm{G}'$, and thus the electric current $j_z$ can control the AFM state through the AFM Edelstein effect induced by the magnetostrictive effect.

Although the current-induced switching is possible in other AFM domains in a similar manner, the magnetically-induced strains in the $xy$ plane are not coupled to the out-ofplane electric current $j_z$. Therefore, the switching of AFM domains having the same toroidal moment cannot be caused by the electric current. This is consistent with the partial toroidic property of the Aizu species. As shown above, by taking the magnetostrictive effect into account, the representation analysis is consistent with the Aizu species analysis in the case of the honeycomb lattice.

\section{Magnetopiezoelectric effect}

In this section, we propose switching of magnetic states by a combination of the electric current and stress. This is similar to switching of structural deformations by making use of the piezoelectric effect, which is the coupling between the electric field and strain. In the following, we introduce a magnetopiezoelectric effect, and discuss the switching of magnetic compounds by using the magnetopiezoelectric effect.

First, let us consider how we manipulate domains of a metallic antiferromagnet U$_3$Ru$_4$Al$_{12}$~\cite{supplemental_Troc2012}. The species of U$_3$Ru$_4$Al$_{12}$ is denoted as 
    \begin{equation}
      \aizu{6/mmm1'}{mmm'\Braket{z}},\label{U3Ru4Al12_Aizu_species}
    \end{equation}
property of which is described as follows [Table III in the main text].
    \begin{center}
    \begin{tabular}{ccccc} \hline \hline 
    \multirow{2}{*}{$\aizu{6/mmm1'}{mmm'\Braket{z}}$}&$\hat{\epsilon}$&$\bm{P}$&$\bm{M}$&$\bm{T}$ \\  
    &P&Z&Z&P\\ \hline\hline
    \end{tabular}
    \end{center}
This implies that none of the physical quantities (strain $\hat{\epsilon}$, electric polarization $\bm{P}$, ferromagnetic moment $\bm{M}$, and toroidal moment $\bm{T}$) distinguish the domain states of U$_3$Ru$_4$Al$_{12}$ in the complete way. The magnetopiezoelectricity derived from asymmetric distortions of the electronic band structure, however, may be useful for the perfect distinction of the AFM domains~\cite{supplemental_Watanabe2017magnetic,supplemental_Watanabe2018grouptheoretical}.

Next, we introduce the magnetopiezoelectric effect. The asymmetric dispersion in the energy spectrum is realized in systems where both of the parity and time-reversal symmetries are broken. The antisymmetric part in the energy dispersion leads to an electronic nematicity under an electric current. For example, in a system with an asymmetric dispersion which is symmetry-adapted to the basis $k_xk_yk_z$, the electric current $j_z$ gives rise to the electronic nematic order in the $k_xk_y$ plane. Accordingly, the electronic nematicity induces an ionic displacement, that is, the strain represented by $\epsilon_{xy}$. The coupling between the electric current and the electronic/ionic nematic order is called magnetopiezoelectricity~\cite{supplemental_Watanabe2017magnetic}, since the time-reversal symmetry breaking is necessary.

The sign of the magnetopiezoelectric effect differs between the AFM domains connected by the operation $\theta$. Thus, the AFM domains of U$_3$Ru$_4$Al$_{12}$ are completely distinguished by the magnetopiezoelectric effect, and the Aizu species can be classified as full magnetopiezoelectric species. In order words, the AFM domains are perfectly manipulated by making use of the electric current and stress. Although the electric current $j_z$ can switch the domains only in an incomplete way as discussed in the main text, the stress can control the magnetostrictive strains and therefore manipulate the domains with the same toroidal moment $\bm{T}\parallel \hat{z}$. Thus, combination of the electric current and stress enables the perfect AFM domain switching.

The full magnetopiezoelectric property is satisfied in most of the zero toroidic species when both of the space-inversion and time-reversal symmetries are broken. Hence, the zero toroidic magnetic compounds may be manipulated by applying the electric current and stress, while domains cannot be inverted by only the electric current  because of an absence of the ferrotoroidic order. It should be noticed that the strain-free (Ga,Mn)As discussed in Sec.~\ref{AppSection_strainedGaMnAs} is zero toroidic but full magnetopiezoelectric, and the ferromagnetic domains can be manipulated by a combined use of the electric current and stress~\cite{supplemental_Chernyshov2009}.

\section{Candidate materials for electrical switching of antiferromagnetic order} \label{app_candidates}

Our symmetry analysis uncovers a lot of candidate materials for electrically switchable antiferromagnets besides CuMnAs and Mn$_2$Au~\cite{supplemental_Wadley2016,supplemental_Bodnar2018}. In  the following, we discuss some candidate materials we identified with a focus on the $\mathcal{PT}$-symmetric and $\bm{Q}=\bm{0}$ magnetic states. At the end of this section, we also show the list of more than 50 candidate materials.

\subsection{ BaMn\texorpdfstring{$_2$}{TEXT}As\texorpdfstring{$_2$}{TEXT} and CeMn\texorpdfstring{$_2$}{TEXT}Ge\texorpdfstring{$_2$}{TEXT}}

BaMn$_2$As$_2$ and CeMn$_2$Si$_2$ crystallize in a tetragonal structure, which is a well-known ThCr$_2$Si$_2$-type structure (space group: I$4/mmm$,~No.~139). Both compounds undergo the AFM phase transitions~\cite{supplemental_Singh2009b,supplemental_Singh2009c,supplemental_MdDin2015} and the corresponding Aizu species are given by  
  \begin{equation}
    \aizu{4/mmm1'}{4'/m'm'm} \textrm{\ ($z$-collinear AFM)},
  \end{equation}
for BaMn$_2$As$_2$, and
  \begin{equation}
    \aizu{4/mmm1'}{mmm' \Braket{x}} \textrm{\ ($x$-, $y$-collinear AFM)},
  \end{equation}
for CeMn$_2$Si$_2$. The former species is zero toroidic, and the AFM state is not controllable by the electric current $\bm{j}$. On the other hand, the latter species is the same as CuMnAs, that is, full toroidic species indicating the AFM state perfectly controllable by $\bm{j}$.

The above classification is supported by the representation analysis. The magnetic sites Mn are characterized by the site-symmetry group $\bm{H}=\bar{4}m2$. The magnetic representation is obtained as
  \begin{equation}
  \Gamma_{\bm{G}}^\textrm{mag} (\bm{H}) 
    = A_{2g}+E_{g}+B_{1u}+E_u. \label{ThCr2Si2_magnetic_representation}
  \end{equation} 
The AFM states of BaMn$_2$As$_2$ and CeMn$_2$Si$_2$ are characterized by the $B_{1u}$ and the $E_u$ irreducible representation, respectively. While the former irreducible representation is nonpolar, the latter is polar and comprises the in-plane toroidal moment $\{T_x, T_y\}$ in its basis. Thus, although the AFM order of BaMn$_2$As$_2$ cannot be controlled by an electric current, the AFM order of CeMn$_2$Si$_2$ is switchable by the in-plane electric current as in the case of CuMnAs. Therefore, CeMn$_2$Si$_2$ and related materials may be a new electrically switchable antiferromagnets.

\subsection{Trigonal \texorpdfstring{$X$}{TEXT}Mn\texorpdfstring{$_2$}{TEXT}\texorpdfstring{\textbf{\textit{Pn}}$_2$}{TEXT}}

Some Mn 122-compounds denoted by the chemical formula \textit{X}Mn$_2$\textit{Pn}$_2$ crystallize in a trigonal structure (space group: P$\bar{3}m1$,~No.~164) which is different from the ThCr$_2$Si$_2$-type structure. A lot of the compounds have the AFM phase where the magnetic moments of Mn atoms are collinear in the $xy$ plane. In the cases of $\left(X,Pn \right)=$(Ca,~Sb)~\cite{supplemental_McNally2015c}, (Sr,~As)~\cite{supplemental_Das2017}, and (Sr,~Sb)~\cite{supplemental_Sangeetha2018}, the species is obtained as
  \begin{equation}
    \aizu{\bar{3}m1'}{2'/m},\label{SrMn2As2_Aizu_species}
  \end{equation} 
which is full toroidic, and hence the AFM state is perfectly controllable by the electric current $\bm{j}$. With the site-symmetry group of Mn sites $\bm{H}=3m$, the magnetic representation is given by
  \begin{equation}
  \Gamma_{\bm{G}}^\textrm{mag} (\bm{H}) 
    = A_{2g}+E_{g}+B_{1u}+E_u. \label{CaAl2Si2_magnetic_representation}
  \end{equation}
The AFM state in the species~\eqref{SrMn2As2_Aizu_species} is characterized by the $E_u$ irreducible representation. This means that the AFM order is perfectly switchable by an in-plane electric current, since the basis of the $E_u$ representation can be taken as the in-plane toroidal moment. Hence, the antiferromagnets belonging to the species~\eqref{SrMn2As2_Aizu_species} may be a platform of the AFM spintronics. Thus, candidate materials are not restricted to the previously-studied tetragonal systems~\cite{supplemental_Wadley2016,supplemental_Bodnar2018}. 

Most of the trigonal Mn 122-compounds are insulating or semiconducting. The switching may not be efficient, since the AFM Edelstein effect is determined by the Fermi-surface term~\cite{supplemental_Zelezny2017,supplemental_Watanabe2017magnetic}. On the other hand, it has been recently reported that EuMn$_2$As$_2$ becomes metallic by doping hole carriers~\cite{supplemental_Anand2016}. Although the magnetic structure of the doped system has not been identified, it may be a candidate for antiferromagnetic spintronics.

The trigonal \textit{X}Mn$_2$\textit{Pn}$_2$ is a good example to illuminate a complementary role of two methods of symmetry analysis we present. In the case of $\left(X,Pn \right)=$(Ca,Bi)~\cite{supplemental_Gibson2015}, magnetic moments in the basal plane are slightly tilted from the basal rotation axes. The species is obtained as 
  \begin{equation}
    \aizu{\bar{3}m1'}{\bar{1}'}.\label{CaMn2Bi2_Aizu_species}
  \end{equation}
Therefore, the species of CaMn$_2$Bi$_2$ differs from that of other compounds~\eqref{SrMn2As2_Aizu_species}, while both magnetic structures are characterized by the $E_u$ irreducible representation. Correspondingly, the domain states are different between the two species~\eqref{SrMn2As2_Aizu_species} and~\eqref{CaMn2Bi2_Aizu_species}. Thus, although the magnetic representation characterizes the order parameter in the crystalline systems, stable domain states in the magnetic phase may not be uniquely determined. The domain states are completely elucidated with the use of the Aizu species analysis. The two symmetry analysis, the representation analysis and the Aizu species analysis, are complementary to each other.

\renewcommand*{\arraystretch}{1.2}
  \begin{longtable}[htbp]{lllllll} 
  \caption{List of candidate materials for electrically switchable antiferromagnet. The table lists compounds, space group, symmetry of magnetic structure denoted by Aizu species and irreducible representations ($\Gamma^\mathrm{mag}$), conducting properties (M/I), N\'eel temperatures ($T_\mathrm{N}$), and references (Ref.). The blank part has not been clarified to the best of our knowledge. The numbers of the Aizu species and the space groups follow Ref.~\cite{supplemental_Litvin2008} and~\cite{supplemental_internatinaltables}, respectively. All the compounds are characterized by the full toroidic species except for a partial toroidic compound U$_3$Ru$_4$Al$_{12}$. In the column of $\Gamma^\mathrm{mag}$, the toroidal moments $\bm{T}$ are also shown as the basis for the polar irreducible representations.}
  \label{app_magnetic_candidates}\\
    Compounds &Space group&Aizu species&$\Gamma^\mathrm{mag}$&M/I&$T_\mathrm{N}$&Ref.  \vspace{.1cm}\\ \hline \hline  
%
%
  $\rm CeMn_2 Ge_2  $&I$4/mmm$ (139)&$\aizu{4/mmm1'}{mmm'\Braket{x}}$ (218)&$E_{u}\left(\{T_x, T_y\}\right)$&&$318<T<417$&\cite{supplemental_MdDin2015}\\[2pt] 
  $\rm CeMnAsO  $&P$4/nmm$ (129)&$\aizu{4/mmm1'}{4'/m'm'm}$ (252)&$B_{1u}$&&$35<T<340$&\cite{supplemental_Zhang2015f}\\[2pt]
  &&$\aizu{4'/m'm'm}{m'mm'\Braket{x}}$ (-)&$E_{u}\left(\{T_x, T_y \}\right)$&&$7<T<35$&\cite{supplemental_Zhang2015f}\\[2pt]
  &&$\aizu{m'mm'\Braket{x}}{2'/m\Braket{z}}$ (-)&$E_{u}\left(\{T_x, T_y \}\right)$&&$7$&\cite{supplemental_Zhang2015f}\\[2pt]
  $\rm CeMnSbO  $&P$4/nmm$ (129)&$\aizu{4/mmm1'}{4'/m'm'm}$ (252)&$B_{1u}$&&$4.5<T<240$&\cite{supplemental_Zhang2016k}\\[2pt]
  &&$\aizu{4'/m'm'm}{m'mm'\Braket{x}}$ (-)&$E_{u}\left(\{T_x, T_y \}\right)$&&$4.5$&\cite{supplemental_Zhang2016k}\\[2pt]
  $\rm PrMnSbO  $&P$4/nmm$ (129)&$\aizu{4/mmm1'}{4'/m'm'm}$ (252)&$B_{1u}$&metal&$35<T<230$&\cite{supplemental_kimber2010local}\\[2pt]
          &&$\aizu{4/mmm1'}{mmm'\Braket{x}}$ (218)&$E_{u}\left(\{T_x, T_y \}\right)$&metal&35&\cite{supplemental_kimber2010local}\\[2pt]
  $\rm NdMnAsO  $&P$4/nmm$ (129)&$\aizu{4/mmm1'}{4'/m'm'm}$ (252)&$B_{1u}$&semiconductor&$23<T<359$&\cite{supplemental_Marcinkova2010,supplemental_Emery2011b}\\[2pt]
  &&$\aizu{4'/m'm'm}{mmm'\Braket{x}}$ (-)&$E_{u}\left(\{T_x, T_y \}\right)$&semiconductor&23&\cite{supplemental_Marcinkova2010,supplemental_Emery2011b}\\[2pt]
  $\rm DyB_4  $&P$4/mbm$ (127)&$\aizu{4/mmm1'}{mmm'\Braket{x}}$ (218)&$E_{u}\left(\{T_x, T_y \}\right)$&metal&$12.7<T<20.3$&{\cite{supplemental_Fisk1981,supplemental_Will1979,supplemental_Ji2007}}\\[2pt]
  $\rm ErB_4  $&P$4/mbm$ (127)&$\aizu{4/mmm1'}{mmm'\Braket{x}}$ (218)&$E_{u}\left(\{T_x, T_y \}\right)$&metal&13&{\cite{supplemental_Fisk1981,supplemental_Will1979,supplemental_Will1981}}\\[2pt]
  $\rm EuTiO_3$&I$4_2/mcm$ (140)&$\aizu{4/mmm1'}{mmm'\Braket{d}}$ (218)&$E_u~\left(\{T_x, T_y \}\right)$&&5.3&\cite{supplemental_Scagnoli2012}\\[2pt]
  $\rm Mn_2 Au$&I$4/mmm$ (139)&$\aizu{4/mmm1'}{mmm'\Braket{d}}$ (218)&$E_{u}\left(\{T_x, T_y \}\right)$&metal&$>1000$&\cite{supplemental_Barthem2013}\\[2pt]
  $\rm Fe Sn_2$&I$4/mcm$ (140)&$\aizu{4/mmm1'}{mmm'\Braket{d}}$ (218)&$E_{u}\left(\{T_x, T_y \}\right)$&metal&$93<T\lesssim378$&\cite{supplemental_Venturini1987,supplemental_Armbruster2010}\\[2pt]
  &&$\aizu{mmm'(s)}{2'/m\Braket{z}}$ (-)&$E_{u}\left(\{T_x, T_y \}\right)$&metal&$93\lesssim T<378$&\cite{supplemental_Venturini1987,supplemental_Armbruster2010}\\[2pt]
  $\rm CuMnAs$&P$4/nmm$ (129)&$\aizu{4/mmm1'}{mmm'\Braket{x}}$ (218)&$E_{u}\left(\{T_x, T_y \}\right)$&semiconductor&480&\cite{supplemental_Wadley2013}\\[2pt]
  $\rm Cr_2 WO_6$&P$4_2/mnm$ (136)&$\aizu{4/mmm1'}{mmm'\Braket{x}}$ (218)&$E_{u}\left(\{T_x, T_y \}\right)$&&45&\cite{supplemental_Kunnmann1968magnetic,supplemental_Zhu2014}\\[2pt]
  $\rm Cr_2 TeO_6$&P$4_2/mnm$ (136)&$\aizu{4/mmm1'}{mmm'\Braket{x}}$ (218)&$E_{u}\left(\{T_x, T_y \}\right)$&&93&\cite{supplemental_Kunnmann1968magnetic,supplemental_Zhu2014}\\[2pt]
%
%
  $\rm U_3 Ru_4 Al_{12} $&P$6_3/mmc$ (194)&$\aizu{6/mmm1'}{mmm'\Braket{z}}$ (481)&$A_{2u}\left(T_z\right), E_{2u}$&metal&9.5&\cite{supplemental_Pasturel2009a,supplemental_Troc2012}\\[2pt]
%
%
  $\rm CaMn_2As_2 $&P$\bar{3}m1$ (164)&&&semiconductor&62&\cite{supplemental_Sangeetha2016}\\[2pt]
  $\rm CaMn_2Sb_2 $&P$\bar{3}m1$ (164)&$\aizu{\bar{3}m1'}{2'/m}$ (295)&$E_{u}\left(\{T_x, T_y \}\right)$&insulator&85&\cite{supplemental_McNally2015c}\\[2pt]
  &&$\aizu{\bar{3}m1'}{\bar{1}'}$ (286)&$A_{1u},E_{u}\left(\{T_x, T_y \}\right)$&&85&\cite{supplemental_Bridges2009}\\[2pt]
  $\rm CaMn_2Bi_2 $&P$\bar{3}m1$ (164)&$\aizu{\bar{3}m1'}{\bar{1}'}$ (286)&$E_{u}\left(\{T_x, T_y \}\right)$&semiconductor&154&\cite{supplemental_Gibson2015}\\[2pt]
  $\rm SrMn_2P_2  $&P$\bar{3}m1$ (164)&&&semiconductor&53&\cite{supplemental_Brock1994}\\[2pt]
  $\rm SrMn_2As_2 $&P$\bar{3}m1$ (164)&$\aizu{\bar{3}m1'}{2'/m}$ (295)&$E_{u}\left(\{T_x, T_y \}\right)$&insulator&118&\cite{supplemental_Sangeetha2016,supplemental_Das2017}\\[2pt]
  $\rm SrMn_2Sb_2 $&P$\bar{3}m1$ (164)&$\aizu{\bar{3}m1'}{2'/m}$ (295)&$E_{u}\left(\{T_x, T_y \}\right)$&semiconductor&110&\cite{supplemental_Sangeetha2018}\\[2pt]
  $\rm EuMn_2As_2 $&P$\bar{3}m1$ (164)&&&semiconductor &142&\cite{supplemental_Anand2016}\\[2pt]
  $\rm YbMn_2Sb_2 $&P$\bar{3}m1$ (164)&$\aizu{\bar{3}m1'}{1'}$ (284)&$A_{1u},E_{u} \left(\{T_x, T_y \}\right)$&&120&\cite{supplemental_Morozkin2006}\\[2pt]
  $\rm Co_4 Nb_2 O_9  $&P$\bar{3}c1$ (165)&$\aizu{\bar{3}m1'}{\bar{3}'m'}$ (313)&$A_{1u} $&insulator&27.4&\cite{supplemental_Bertaut1961}\\[2pt]
    &&$\aizu{\bar{3}m1'}{2/m'}$ (296)&$E_{u}\left(\{T_x, T_y \}\right)$&insulator&27.2&\cite{supplemental_Khanh2016a}\\[2pt]
  $\rm MnTiO_3  $&R$\bar{3}$ (148)&$\aizu{\bar{3}1'}{\bar{3}'}$ (264)&$A_{u}\left(T_z\right)$&insulator&64&\cite{supplemental_Silverstein2016,supplemental_Shirane1959a}\\[2pt]
  $\rm MnGeO_3  $&R$\bar{3}$ (148)&$\aizu{\bar{3}1'}{\bar{3}'}$ (264)&$A_{u}\left(T_z\right)$&&120&\cite{supplemental_Tsuzuki1974c}\\[2pt]
%
%
  $\rm NdCrTiO_5  $&Pbam (55)&$\aizu{mmm1'}{mmm'}$ (71)&$B_{1u}\left(T_z\right)$&insulator&13&\cite{supplemental_Buisson1970}\\[2pt]
  &&&&&21&\cite{supplemental_Hwang2012}\\[2pt]
  $\rm LiFePO_4 $&P$nma$ (62)&$\aizu{mmm1'}{mmm'}$ (71)&$ B_{1u} \left(T_z\right)$&insulator&50&\cite{supplemental_Santoro1967} \\[2pt]
  &&$\aizu{mmm1'}{2/m'\Braket{z}}$ (60)&$ A_{u}, B_{1u} \left(T_z\right)$&&47&\cite{supplemental_Li2006,supplemental_Toft-Petersen2015} \\[2pt]
  $\rm LiNiPO_4 $&P$nma$ (62)&$\aizu{mmm1'}{mmm'}$ (71)&$ B_{2u} \left(T_y\right)$&insulator&20.8&\cite{supplemental_Kornev2000a} \\[2pt]
  $\rm LiCoPO_4 $&P$nma$ (62)&$\aizu{mmm1'}{mmm'}$ (71)&$ B_{1u} \left(T_z\right)$&insulator&21.6&\cite{supplemental_Fogh2017,supplemental_santoro1966magnetic}\\[2pt]
    &&$\aizu{mmm1'}{2'\Braket{x}}$ (53)&$ B_{2g},B_{1u} \left(T_z\right),B_{2u} \left(T_y\right) $&&&\cite{supplemental_Vaknin2002,supplemental_VanAken2007a}\\[2pt]
  $\rm KMn_4 (PO_4)_3 $&P$nam$ (62)&$\aizu{mmm1'}{mmm'}$ (73)&$ B_{2u} \left(T_y\right)$&&$10$&\cite{supplemental_Lopez2008}\\[2pt]
  $\rm t-NaFePO_4 $&P$nma$ (62)&$\aizu{mmm1'}{mmm'}$ (71)&$ B_{1u} \left(T_z\right) $&&50&\cite{supplemental_Avdeev2013} \\[2pt]
  $\rm Gd_5 Ge_4  $&P$nma$ (62)&$\aizu{mmm1'}{mmm'}$ (71)&$ B_{1u} \left(T_z\right)$&metal&127&\cite{supplemental_Tan2005,supplemental_Levin2001} \\[2pt]
  $\rm EuZrO_3  $&P$nma$ (62)&$\aizu{mmm1'}{mm'm}$ (71)&$B_{2u} \left(T_y\right)$&insulator&4.1&\cite{supplemental_Avdeev2014}\\[2pt]
  &&$\aizu{mmm1'}{m'm'm'}$ (73)&$A_{u}$&insulator&4.4&\cite{supplemental_Saha2016c}\\[2pt]
  $\rm TbCoO_3  $&P$bnm$ (62)&$\aizu{mmm1'}{mmm'}$ (71)&$B_{1u} \left(T_z\right)$&insulator&3.31&\cite{supplemental_Knizek2014}\\[2pt]
  $\rm HoCoO_3  $&P$nma$ (62)&$\aizu{mmm1'}{m'mm'}$ (71)&$B_{2u} \left(T_y\right)$&&3&\cite{supplemental_Munoz2012}\\[2pt]
  $\rm MnNb_2O_6  $&P$bcn$ (60)&$\aizu{mmm1'}{2'/m\Braket{x}}$ (59)&$B_{2u} \left(T_y\right),B_{3u} \left(T_x\right)$&&4.4&\cite{supplemental_Jacobson1975}\\[2pt]
  $\rm CoSe_2O_5  $&P$bcn$ (60)&$\aizu{mmm1'}{m'mm}$ (71)&$B_{3u} \left(T_x\right)$&&8.5&\cite{supplemental_Melot2010b}\\[2pt]
  $\rm TbGe_2 $&C$mmm$ (65)&$\aizu{mmm1'}{m'mm}$ (71)&$B_{3u} \left(T_x\right)$&&41&\cite{supplemental_Schobinger-Papamantellos1988}\\[2pt]
  $\rm Ce_3Sn_7 $&C$mmm$ (65)&$\aizu{mmm1'}{m'mm}$ (71)&$B_{3u} \left(T_x\right)$&metal&5&\cite{supplemental_Bonnet1994,supplemental_Givord1989}\\[2pt]
  $\rm Sm_3 Ag_4 Sn_4 $&I$mmm$ (71)&$\aizu{mmm1'}{mmm'}$ (71)&$B_{1u}\left(T_z\right)$&&8.3&\cite{supplemental_Voyer2007}\\[2pt]
  &&$\aizu{mmm1'}{mm'm}$ (71)&$B_{2u}\left(T_y\right)$&&8.3&\cite{supplemental_Voyer2007}\\[2pt]
  $\rm UCu_5In  $&P$nma$ (62)&$\aizu{mmm1'}{mm'm}$ (71)&$B_{2u}\left(T_y\right)$&metal&25&\cite{supplemental_Tran2001}\\[2pt]
  $\rm KFeO_2 $&P$bca$ (61)&$\aizu{mmm1'}{m'm'm'}$ (73)&$A_{u}$&&960&\cite{supplemental_Tomkowicz1977a}\\[2pt]
  &&$\aizu{mmm1'}{m'mm}$ (71)&$B_{3u} \left(T_x\right)$&&$\sim 1001$&\cite{supplemental_Sheptyakov2010}\\[2pt]
  $\rm CoGeO_3  $&P$bca$ (61)&$\aizu{mmm1'}{mmm'}$ (71)&$B_{1u}\left(T_z\right)$&insulator&33.1&\cite{supplemental_Redhammer2010a}\\[2pt]
  $\rm DyVO_4 $&I$mma$ (74)&$\aizu{mmm1'}{mmm'}$ (71)&$B_{1u}\left(T_z\right)$&insulator&3.8&\cite{supplemental_Will1971,supplemental_Kishimoto2010}\\[2pt]
  YbAl$_{1-x}$Fe$_x$B$_{4}$&P$bam$ (55)&$\aizu{mmm1'}{mmm'}$ (71)&$B_{3u} \left(T_x\right)$&metal&&\cite{supplemental_ybalb4}\\[2pt]
              &P$bam$ (55)&$\aizu{mmm1'}{m'm'm'}$ (73)&$A_{u}$&metal&&\cite{supplemental_ybalb4}\\[2pt]
%
%
  $\rm Co_3 Te O_6  $&C$2/c$ (15)&$\aizu{2/m1'}{2'/m}$ (26)&$B_{u}\left(T_x, T_y\right)$&&21.1&\cite{supplemental_Ivanov2012}\\[2pt]
  $\rm MnPS_3 $&C$2/m$ (12)&$\aizu{2/m1'}{2'/m}$ (26)&$B_{u}\left(T_x, T_y\right)$&insulator&78&\cite{supplemental_Kurosawa1983,supplemental_Ressouche2010a}\\[2pt] 
  $\rm LiFeSi_2O_6  $&P$2_1/c$ (14)&$\aizu{2/m1'}{2/m'}$ (27)&$ A_{u}\left(T_z\right) $& &17.8&\cite{supplemental_Redhammer2009b,supplemental_Redhammer2001} \\[2pt]
  &&$\aizu{2/m1'}{\bar{1}'}$ (17)&$ A_{u}\left(T_z\right),B_u \left(T_x, T_y\right) $& &18&\cite{supplemental_Toledano2015} \\[2pt]
  $\rm LiCrSi_2O_6  $&P$2_1/c$ (14)&$\aizu{2/m1'}{2'/m}$ (26)&$ B_{u}\left(T_x, T_y\right) $&&11.5&\cite{supplemental_Nenert2010} \\[2pt]
  $\rm LiCrGe_2O_6  $&P$2_1/c$ (14)&$\aizu{2/m1'}{2'/m}$ (26)&$ B_{u}\left(T_x, T_y\right) $&&4.8&\cite{supplemental_Nenert2010,supplemental_Nenert2009a} \\[2pt]
  $\rm LiVGe_2O_6 $&P$2_1/c$ (14)& &$A_u \left(T_z\right)$ or $B_{u}\left(T_x, T_y\right) $&&24&\cite{supplemental_Lumsden2000} \\[2pt]
  $\rm NaCrSi_2O_6  $&C$2/c$ (15)&$\aizu{2/m1'}{\bar{1}'}$ (17)&$ A_u \left(T_z\right), B_{u}\left(T_x, T_y\right)$&&2.8&\cite{supplemental_Nenert2010a} \\[2pt]
  $\rm CaMnGe_2O_6  $&C$2/c$ (15)&$\aizu{2/m1'}{\bar{1}'}$ (17)&$ A_{u}\left(T_z\right), B_u $&&12&\cite{supplemental_Redhammer2008} \\[2pt]
  &&$\aizu{2/m1'}{2'/m}$ (26)&$ B_u \left(T_x , T_y\right) $&insulator&15&\cite{supplemental_Ding2016} \\[2pt]
  $\rm MnGeO_3  $&C$2/c$ (15)&$\aizu{2/m1'}{2'/m}$ (26)&$ B_{u}\left(T_x, T_y\right) $&&35.1&\cite{supplemental_Redhammer2011} \\[2pt]

  $\rm Na_2RuO_4$&P$2_1/c$ (14)&$\aizu{2/m1'}{2/m'}$ (27)&$A_{u}\left(T_z\right)$&&37.22&\cite{supplemental_Mogare2006}\\
  \renewcommand{\arraystretch}{1}
  \end{longtable}

\pagebreak
\input{supple_refs}
\end{document}

%% file: supple_refs.tex
%